\documentclass[twocolumn]{aastex62}
\pdfoutput=1
\date{\today}
\usepackage{graphicx}
\usepackage[utf8]{inputenc}
\usepackage{amsmath,amstext}
\usepackage[T1]{fontenc}
\usepackage{apjfonts} 
\usepackage[figure,figure*]{hypcap}
\usepackage{amssymb}
\usepackage{color}
\usepackage{hyperref}
\usepackage{longtable}
\newlength\figureheight
\newlength\figurewidth
\usepackage[caption=false]{subfig}

\usepackage[amssymb,mediumspace,thickqspace]{SIunits}

\newcolumntype{G}[0]{>{\displaystyle }l<{}}

\shorttitle{The hot dust of ESO323-G77}
\shortauthors{Leftley J.H. et al.}

\begin{document}
\author[0000-0001-6009-1803]{James H. Leftley}
\affiliation{Department of Physics \& Astronomy, University of Southampton, Southampton, SO17 1BJ, UK}
\affiliation{Universit\'e de la C\^ote d’Azur, Observatoire de la C\^ote d’Azur, CNRS, Laboratoire Lagrange, Parc Valrose, B\^at. H. Fizeau, 06108 Nice, France}
\affiliation{European Southern Observatory, Alonso de Córdova 3107, Casilla 19001, Santiago, Chile}

\author[0000-0001-8281-5059]{Konrad R. W. Tristram}
\affiliation{European Southern Observatory, Alonso de Córdova 3107, Casilla 19001, Santiago, Chile}

\author[0000-0002-6353-1111]{Sebastian F. H\"onig}
\affiliation{Department of Physics \& Astronomy, University of Southampton, Southampton, SO17 1BJ, UK}

\author[0000-0003-0220-2063]{Daniel Asmus}
\affiliation{Department of Physics \& Astronomy, University of Southampton, Southampton, SO17 1BJ, UK}

\author[0000-0002-2216-3252]{Makoto Kishimoto}
\affiliation{Department of Astrophysics and Atmospheric Sciences, Kamigamo-Motoyama, Kita-ku, Kyoto, 603-8555, Japan}


\author[0000-0003-3105-2615]{Poshak Gandhi}
\affiliation{Department of Physics \& Astronomy, University of Southampton, Southampton, SO17 1BJ, UK}

\title{Resolving the Hot Dust Disk of ESO323-G77}

\begin{abstract}
Infrared interferometry has fuelled a paradigm shift in our understanding of the dusty structure in the central parsecs of Active Galactic Nuclei (AGN). The dust is now thought to comprise of a hot ($\sim1000\,$K) equatorial disk, some of which is blown into a cooler ($\sim300\,$K) polar dusty wind by radiation pressure. In this paper, we utilise the new near-IR interferometer GRAVITY on the Very Large Telescope Interferometer (VLTI) to study a Type 1.2 AGN hosted in the nearby Seyfert galaxy ESO\,323-G77. By modelling the squared visibility and closure phase, we find that the hot dust is equatorially extended, consistent with the idea of a disk, and shows signs of asymmetry in the same direction. Furthermore, the data is fully consistent with the hot dust size determined by K band reverberation mapping as well as the predicted size from a \textit{CAT3D-WIND} model created in previous work using the SED of ESO\,323-G77 and observations in the mid-IR from VLTI/MIDI.
\end{abstract}

\section{Introduction}

IR interferometry has played an integral part in the development of our understanding of Active Galactic Nuclei (AGN). Observations with the Mid-Infrared Interferometer \citep[MIDI,][]{leinert_midi_2003} at the Very Large Telescope Interferometer (VLTI) allowed the study of the central tens of parsecs in the mid-IR. At this scale, it was thought that a dusty toroidal structure, believed to be the source of obscuration required for the difference between Type 1 and Type 2 AGN in the unification scheme, would be found \citep[e.g.][]{antonucci_unified_1993}. What was found instead was a polar dust structure \citep[e.g.][]{tristram_resolving_2007,kishimoto_mapping_2011,honig_parsec-scale_2012,honig_dust_2013,burtscher_diversity_2013,tristram_dusty_2014,honig_dust-parallax_2014,lopez-gonzaga_mid-infrared_2016}.

The discovery of the polar dust structure spurred the development of the disk+wind model \citep{emmering_magnetic_1992,elitzur_agn-obscuring_2006,honig_dust_2013}. In the model, the polar dust is an outflow that must have an origin where it receives material and energy. The source of the material in the disk+wind model is an equatorial hot dust disk that is close to the central engine of the AGN \citep{honig_dusty_2017}. Such a disk would be on scales too small for MIDI to study in most AGN and is thought to be responsible for the unresolved emission observed with MIDI. The disk geometry was drawn from the warm dust disk seen with MIDI in the nearby AGN hosted by the Circinus Galaxy \citep{tristram_complexity_2012,tristram_dusty_2014}.

The source of obscuration in the disk+wind scenario, necessary for AGN unification, is from the hot dust disk and the launch region of the dusty wind. The cool extended wind could also explain the apparent isotropic mid-IR appearance of AGN, irrespective of inferred line-of-sight obscuration \citep{gandhi_resolving_2009, asmus_subarcsecond_2015, honig_dusty_2017}. The hot dust should primarily emit in the near-IR which makes the near-IR an important regime to test the disk+wind model. So far, AGN have been less studied with near-IR interferometry than with mid-IR interferometry. \citet{weigelt_vlti/amber_2012} studied NGC\,3783 using AMBER \citep[Astronomical Multi-BEam combineR,][]{petrov_amber_2007}; they report a hot, 1400K, thin dust ring fit. \citet{kishimoto_exploring_2009,kishimoto_innermost_2011} reported that near-IR interferometry was likely tracing the sublimation radius of the AGN using Keck observations which agrees with the discovered thin ring/disk. The disk component can be explained in the disk+wind model as the hot dust disk, thought to be responsible for the bulk of the near-IR emission. However, the ring like emission can also be explained in the classic torus model as the expected sublimation radius inside the torus. Further study of the near-IR dust distribution is needed to distinguish the two.

The new instrument GRAVITY at the VLTI presents us with the opportunity to constrain this hot dust disk. It operates in the near-IR, where such a disk would be brightest, and offers access to the spatial scales required. This had been attempted in a few objects already \citep[][]{gravity_collaboration_resolved_2020,gravity_collaboration_image_2020} and structure on the spatial scale of the hot dust disk has been detected. In this work we will attempt to constrain the hot ($\sim1000$\,K) dust in the Type 1.2 nuclei of the local Seyfert galaxy ESO\,323-G77. This object has shown a polar extension in the mid-IR as well as a particularly dominant unresolved component when studied with MIDI \citep{leftley_new_2018}. The unresolved component is thought to be the putative disk in this object which makes it ideal to try and constrain the launch region of the dusty wind.

This paper is structured as follows. In Section\,\ref{Sec:Obs}, we detail the GRAVITY observations and the method of data reduction.  Section\,\ref{Section:ObRes} presents and discusses the results drawn directly from the observations as well as simplistic modelling of the squared visibility. Section\,\ref{sec:Grav_Mod} gives the models used in our attempt to recreate the squared visibility and closure phase. Section\,\ref{Sec:ModRes} presents the results of the modelling and discusses the reliability and interpretation of these results.

\section{Observations and Reduction}\label{Sec:Obs}
We obtained data from the GRAVITY instrument \citep[][]{gravity_collaboration_first_2017}. In total, there was 8 hours of scheduled GRAVITY time split into 4 observations. We used a non-standard observation strategy to compensate for any biases that were introduced from the Adaptive Optics (AO) systems on the VLTI. The bias from AO was discovered when observing faint objects with AMBER. It was found that the poor performance of the AO systems, on the UTs at the VLT, when observing R band faint targets, such as AGN, led to visibility losses (see also \citealt{burtscher_observing_2012} for the effect of poor AO correction in the mid-IR). In principle, this loss can be corrected with a faint red calibrator observed temporally and spatially close to the science target. The faint calibrator is a star that has similar R band properties to the science target. The star would then be calibrated with the normal bright calibrator and any visibility loss present could be quantified corrected in the science.

\begin{figure}
    \centering
    \includegraphics[width=0.45\textwidth,trim={2.6cm 0 1cm 0},clip]{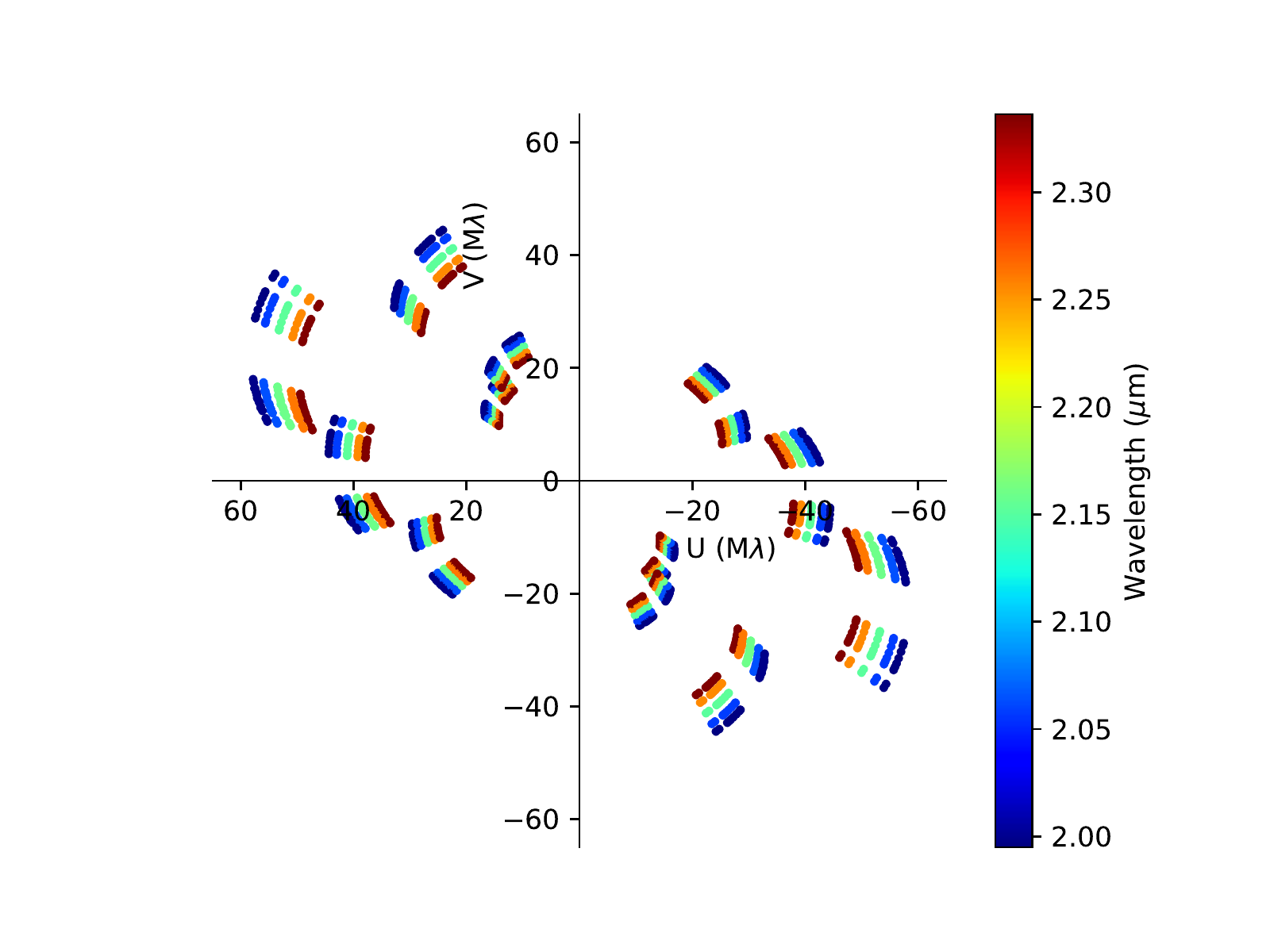}
    \caption{The GRAVITY \textit{uv} plane for ESO\,323-G77.}
    \label{fig:GravUV}
\end{figure}

The observations are listed in Table\,\ref{tab:gravobs}. Each observation block was structured RCAL-SCI-RCAL-CAL, where RCAL is the red calibrator. The SCI component of the April 2019 and May 2019 block is split into three 300\,s observations, the June 2019 block was instead split into six 160\,s observations. Like the June block, the March 2020 block was due to be split into six parts; however, due to poor atmospheric conditions on the night, some observations were repeated and eleven total observations were obtained. Due to the poor conditions, observation 2020-03-15 08:04:37 was unusable and discarded. The UT3-UT2 and UT3-UT1 visibility squared of 2020-03-15 07:34:28 were discarded for the same reason as well as the closure phase for this observation. When performing the data reduction, all observations were reduced and calibrated individually. For the June 2019 night, observations were then paired together in time, and averaged into three observations. For the March 2020 night, the ten sets of visibility squared, in time order, were collected into groups of three-three-two-two and averaged to produce four observations; the nine sets of closure phases were collected into three groups of three and averaged. The groupings are given in Table\,\ref{tab:gravobs}. For every observation, the CAL is HD\,120271, the first RCAL is UCAC4\,237-064201, and the second RCAL is NOMAD1\,0480-0335664. The achieved \textit{uv} plane for the observations of ESO\,323-G77 can be found in Figure\,\ref{fig:GravUV}.

\begin{table}[b]
    \centering
    \begin{tabular}{llccc}
        Date&Time&Obs Length (s)&V Group& C Group \\\hline\hline
        2019-04-23&03:04:24&300\\
        2019-04-23&03:09:59&300\\
        2019-04-23&03:25:07&300\\
        2019-05-13&03:51:06&300\\
        2019-05-13&04:00:33&300\\
        2019-05-13&04:11:48&300\\
        2019-06-20&23:38:04&160&1&1\\
        2019-06-20&23:41:16&160&1&1\\
        2019-06-20&23:47:47&160&2&2\\
        2019-06-20&23:54:06&160&2&2\\
        2019-06-21&00:00:40&160&3&3\\
        2019-06-21&00:03:48&160&3&3\\
        2020-03-15&07:31:13&160&4&4\\
        2020-03-15&07:34:28&160&4&*\\
        2020-03-15&07:41:07&160&4&4\\
        2020-03-15&07:44:19&160&5&4\\
        2020-03-15&07:47:31&160&5&5\\
        2020-03-15&07:54:13&160&5&5\\
        2020-03-15&07:57:28&160&6&5\\
        2020-03-15&08:01:25&160&6&6\\
        2020-03-15&08:04:37&160&*&*\\
        2020-03-15&08:14:04&160&7&6\\
        2020-03-15&08:17:31&160&7&6
    \end{tabular}
    \caption{The GRAVITY observations used in this chapter and their total observation length. If a group is given, objects of the same group were combined when used. V Group is the grouping used when determining the visibility squared, C Group is the grouping used when determining the closure phase. *) This dataset was not included in the analysis.}
    \label{tab:gravobs}
\end{table}{}

The nucleus of ESO\,323-G77 is dim by the standards of the GRAVITY fringe tracker and there is no off-axis bright object with which to perform dual-field observations. When attempted in single field mode, 50\% of the light was not enough to successfully track fringes. However, the fringe tracker data can be used for more than just tracking. Because we are interested in the continuum more than line emission, we did not require high spectral resolution. The fringe tracker itself provides low spectral resolution (R$\sim20$) data, albeit for short integration times. Therefore, we did not need the science channel at all. Instead, 100\% of the target light can be sent to the fringe tracker using the dual-field off-axis mode. The target for the fringe tracker was set to be the nucleus of ESO\,323-G77 and the science target for the science channel was set to be an arbitrary piece of the sky far enough away. DIT = 10\,s for the science combiner was chosen for the 160\,s exposures in order to allow for a full cycle of OPD offsets with the Fringe Tracker to de-bias the data \citep{lacour_gravity_2019}. DIT = 30\,s for the science combiner was chosen for the 300\,s exposures. For the data reduction and the analysis, the science channel was ignored.

\subsection{GRAVITY data reduction}

We make use of the \href{https://www.eso.org/sci/software/cpl/esorex.html}{\textsc{esorex}} pipeline through the \href{https://version-lesia.obspm.fr/repos/DRS_gravity/python_tools/}{python tools} provided by the Gravity Consortium to perform the initial reduction. We calculate the final visibility and phase from the reduced products manually.

To reduce the data, we first use the run\_gravi\_reduce script from the python tools, which utilises \textsc{esorex} with the gravity\_vis recipe. The tool produces intermediary products (given by setting the parameter --p2vmreduced-file=TRUE) which are a set of files that contain the reduced data before post processing as well as some extra relevant information such as the group delay. Most importantly for our reduction, the intermediary products contain the complex visibility for each frame, a rejection flag for each frame, and a geometric flux for each frame. We compute the visibility and the closure phase separately from this point.

\subsubsection{Visibility determination}
To determine the visibility and visibility squared we initially used the default pipeline frame rejection without any smoothing over frames. This uses a signal to noise ratio cut off of 3. We then compared the visibility of all calibrators to the strehl ratio, calculated from the acquisition camera images by the pipeline using the --reduce-acq-cam option in the gravity\_vis \textsc{esorex} recipe. The strehl ratio is our best proxy for the performance of the AO systems. Also considered was the header provided coherence time, the FWHM of a Gaussian fitted to the object as seen by acquisition camera, the core width and power index of a Moffat function fitted to the object, and the variance in central position of the source as determined by the fitted Moffat function, i.e. the "jitter", in the acquisition camera over the course of the observation. All these measures correlated with loss in visibility and visibility squared from AO performance; the strongest correlation was with the strehl ratio and the FWHM.

By modelling this loss, we could remove the majority of the visibility loss from the science exposures but a significant scatter, approximately 0.2 in visibility squared on the best night, still rendered most of the visibility data useless. Therefore we derived a separate method for accounting for this loss.

\begin{figure}
    \centering
    \includegraphics[width=0.49\textwidth,trim={0cm 0 0 0},clip]{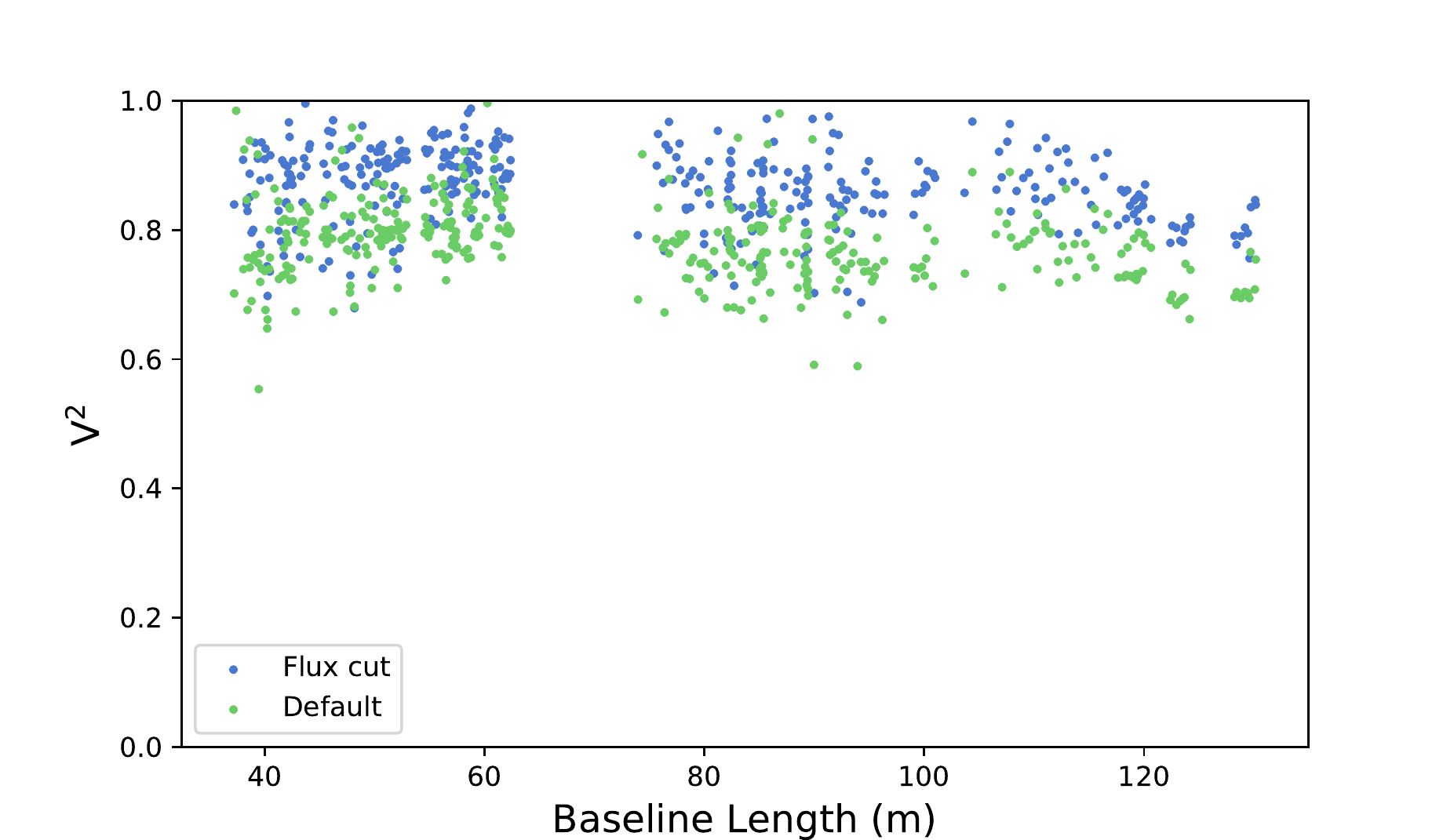}
    \caption{The visibility squared observations, at 2.15$\,\mu$m, 2.26$\,\mu$m, and 2.34$\,\mu$m, using the default pipeline values and the 3\% flux cut. The baseline lengths for each wavelength are adjusted to 2.15$\,\mu$m equivalent. Both are calibrated using the same calibrators.}
    \label{fig:viscutcomp}
\end{figure}

We utilised the individual frame complex visibility and selected frames based on our own criteria. We attempted to use the group delay to select frames, as done in \citet{gravity_collaboration_resolved_2020}; however, we found that selecting by geometric flux provided the best correction in our data. We selected the 3\% brightest frames in geometric flux, not including any frames rejected by the gravity\_vis recipe. All non-selected frames were then flagged and the visibility and visibility squared were calculated using the GRAVITY python tool run\_gravi\_reduce\_from\_p2vmred with the \textsc{esorex} recipe gravity\_vis\_from\_p2vmred using the flag --use-existing-rejection=TRUE. This removed some, but not all, of the AO loss. To remove the remaining bias, we calibrated the science data with the red calibrators. We show the visibility squared data with and without this cut in Figure\,\ref{fig:viscutcomp}. Both the science and calibrators were reduced by the same method.

It is suspected that the reason the AO loss can be removed this way is that when the AO is performing well more flux is injected into the fibre. Therefore, by selecting the highest geometric fluxes, we select the frames where the AO is performing best or where the atmosphere was least turbulent. This method is reminiscent of "lucky imaging" sometimes used for ground based photometry \citep{fried_probability_1978}. The disadvantage of this method is we discard most of the data.

Finally, it is possible that the visibility squared at zero baseline length is not 1. This scaling issue is possibly caused by coherence loss and effects all baselines of the same observation equally. For a single night of observations, where the instrument setup remains constant, this scaling should be negligibly variant between similar \textit{uv} positions \citep{gravity_collaboration_resolved_2020} and can be accounted for when modelling. However, between nights the scaling could change. This scaling does not appear to be completely removed through calibration and, therefore, may be caused by an instrumental response to dimmer K-band objects. The calibrators, including the red calibrators, are brighter in K than the science. To account for the scale variation, we take the median visibility squared value of all science observations during a night and scale it to the night of highest median visibility squared at a similar position angle. This can only be done for observations with similar position angles and baseline lengths because any scaling would be degenerate with angular dependant structure. This equates to the median visibility squared of the March and June nights being scaled to the median visibility squared of the April and May nights respectively. While we cannot account for difference in scaling at different position angles, by normalising to the highest visibility squared nights the different scaling between the different position angles is minimised. The remaining average scaling is then included in the models when fitting.

\subsubsection{Closure phase determination}

Accurate determination of the closure phase important to determine asymmetric detail in the geometry of ESO\,323-G77. While it is piston-invariant there are still some residual offsets that need correcting after the initial pipeline reduction.

To calculate the closure from a single triplet of cotemporal frames in the p2vmred file, we used a similar method to the \textsc{esorex} pipeline. We calculate the complex bispectrum using:
\begin{equation}
    \Psi_{ijk}=\psi_{ij}\cdot\psi_{jk}\cdot\psi^*_{ik}.
\end{equation}{}
where i,j, and k are three unique telescopes and $\psi$ is the complex visibility of a telescope pair. We proceed to follow the method set by the fringe tracker \citep{lacour_gravity_2019}. When observing, the fringe tracker data are averaged over 300 frames when producing the closure phase estimates. Therefore, we split the data into 300 frame segments. In each 300 frame segment, we remove any trio of frames that have any of their three frames flagged with the default pipeline flagging. We do not make any flux cuts for the closure phase determination. We then calculate the mean bispectrum of the segment. We calculate the final closure phase from the angle of the mean of the bispectrum of every segment in the observation (Equation\,\ref{eq:sumofbispectrum}) and the standard error from their distribution.

\begin{equation}\label{eq:sumofbispectrum}
    \Phi_{ijk}=\arg\left(\frac{1}{n}\sum^n_{N=0} \frac{1}{300}\sum^{299+300N}_{f=300N} \Psi_{ijkf}\right).
\end{equation}{}

We calculate this for every science and calibrator observation and find that all calibrators have non-zero closure phase. The bright calibrator has a very large phase of $\approx 7^\circ$ on every night. We do not know what causes this and more bright calibrators would need to be checked to see if this is instrumental or if HD\,120271 is a bad calibrator. The closure phase of the red calibrators on the same night are non-zero and approximately equal; furthermore, the common phase is clearly present in the science. As multiple red calibrators show the same phase component we can conclude this is instrumental. We therefore subtract the mean closure phase of the red calibrators from the science target. The remaining closure phase was taken to be truly from the science object. 

\section{Observational Results and Discussion}\label{Section:ObRes}

After reduction and calibration of the visibility squared (shown in Figure\,\ref{fig:ReverbComp} for 2.15\,$\mu$m) and closure phase, we can already draw some conclusions. There is a clear drop in visibility squared of approximately 0.1 from the 60\,m to 130\,m baseline at all covered wavelengths except for the 1.99$\,\mu$m bin. The shortest baseline (UT3-UT2) has a lower visibility squared than the second shortest baseline in the May, June, and March observations at all observed wavelengths. The lower squared visibility could suggest that the shortest baseline suffers an extra component of loss due to an unknown reason that was not mirrored in the calibrator. The suppressed first baseline is not present in the April observation which is at a similar PA to the June observation, suggesting it is not a real structure. Similar losses are seen in other objects in \citet{gravity_collaboration_resolved_2020}. Therefore, we do not consider the UT3-UT2 baseline when fitting squared visibility.

The first wavelength bin, 1.99$\,\mu$m, is heavily suppressed in visibility and visibility squared by $\sim0.5$ after the 3\% flux cut. It also has a large, $\sim10^\circ$, scattered closure phase. Furthermore, both the visibility and closure phase seem disordered with baseline length when compared to all other wavelength bins. Without the 3\% flux cut, the visibility and visibility squared are often negative. The reason is most likely the contamination from back-scattered light from the GRAVITY metrology laser at 1.908\,$\mu$m \citep[c.f.][]{gravity_collaboration_first_2017}. This contamination has an intensity equivalent to a $\sim 10\,$mag source, determined from the sky observations, for the setup we used, i.e.\ its contribution is comparable to the flux from our science target. We therefore do not consider results from the shortest wavelength bin to be reliable.

The second bin (2.06$\,\mu$m) also has squared visibilities that are $\sim0.1$ lower than the latter three wavelength bins. By itself it can be considered reliable because it shows a similar structure to the latter bins but scaled and the scale of the squared visibilities is separately considered in fitting. As a precaution, we do not use the 2.06$\,\mu$m bin either when performing multi-wavelength fitting so as not to introduce any erroneous structure. Over the last three wavelength bins (2.15$\,\mu$m, 2.26$\,\mu$m, and 2.34$\,\mu$m), the visibility squared has a standard deviation of at most 0.05 at similar baseline lengths. Therefore, the latter three bins can be used together in multi-wavelength fitting.

\begin{figure*}
    \centering
    \includegraphics[width=0.9\textwidth]{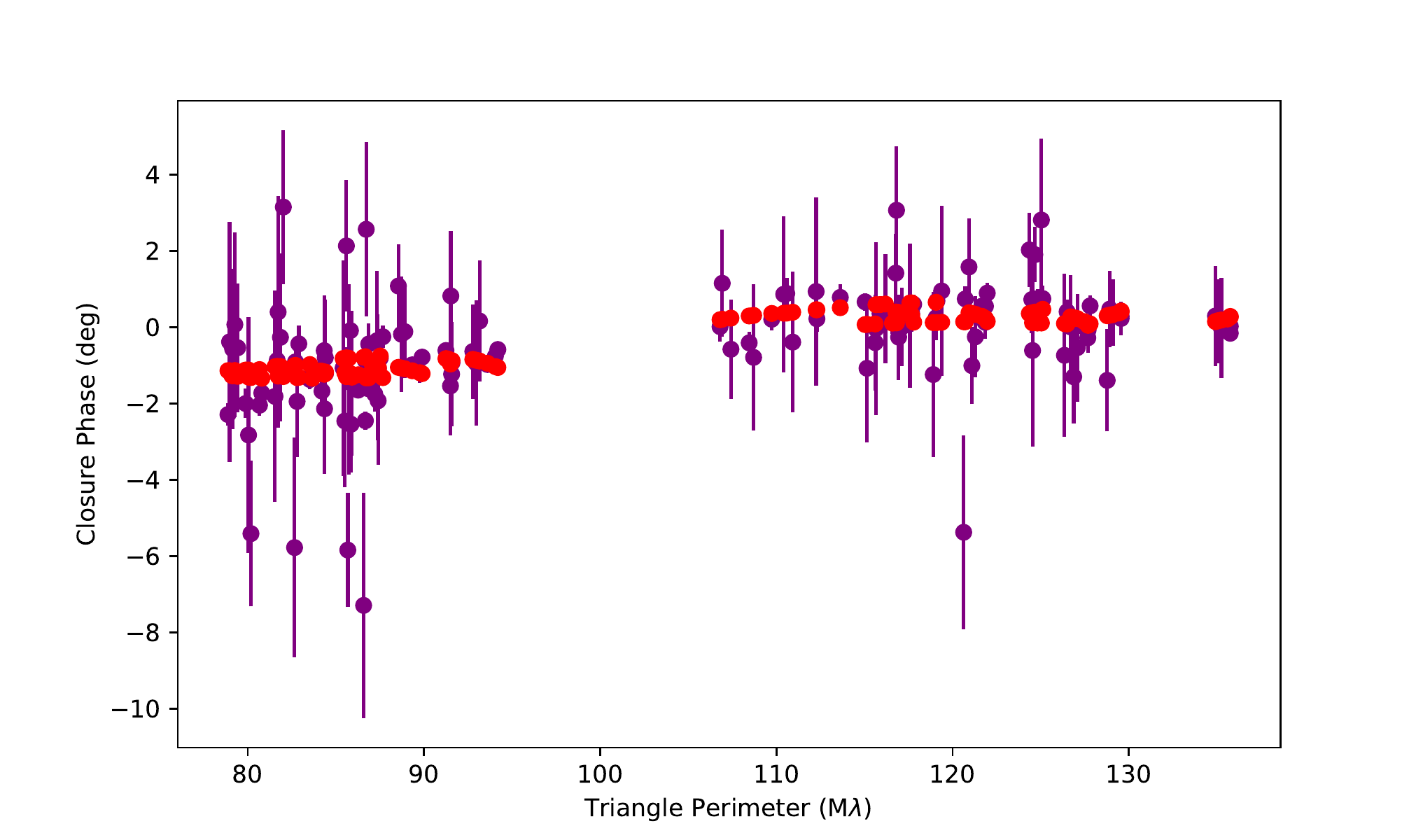}
    \caption{The closure phase for the 2.15$\,\mu$m, 2.26$\,\mu$m, and 2.34$\,\mu$m wavelength bins are plotted in purple against baseline length in M$\lambda$. The red depicts the best fitting model at sampled at the same \textit{uv} locations.}
    \label{fig:GravClosMod}
\end{figure*}

The closure phase in each reliable wavelength bin shows structure. The phase for the 2.15$\,\mu$m, 2.26$\,\mu$m, and 2.34$\,\mu$m wavelength bins is plotted in Figure\,\ref{fig:GravClosMod} in purple. We find no phase for the larger triangles, as measured by the sum of the three baselines of the telescope triplet. The phase gradually decreases to $\sim-1.5^\circ$ for smaller triangles. A small change in phase could be interpreted as a small shift of photocentre at larger spacial scales. The component responsible for the phase would have to be resolved out for the higher spatial resolution triplets.

\subsection{Comparison to reverberation mapping}\label{Sec:Reverb}

As an initial test for structure, we make the assumption that the visibility squared in a single wavelength bin can be explained by a simple 1D Gaussian model. Additionally included in the model is a scale factor; this factor scales all the visibility squared measurements so that the zero baseline value may be different from one due to either visibility loss or an over-resolved structure. The model is as follows:

\begin{equation}\label{eq:GaussAny}
\begin{array}{GGG}
V^2(u,v,\lambda) &=& (1-s_f)*F(u,v,\lambda,\Theta)^2,
\end{array}
\end{equation}
\begin{center}
    where
\end{center}{}
\begin{equation}
\begin{array}{GGG}
F(u,v,\lambda,\Theta)&=&\exp\left[-\left(\frac{ C \Theta }{\lambda }\right)^2 (u^2+v^2)\right],
\end{array}
\end{equation}
\begin{center}
    and
\end{center}{}
\begin{equation}
\begin{array}{GGG}
C&=&\frac{\pi^2}{1.296\times10^9\cdot\,\sqrt[]{\ln2}},
\end{array}
\end{equation}
where $\Theta$ is the FWHM value of the Gaussian in mas, $\lambda$ is the wavelength, and $s_f$ is the scale factor. The constant $C$ comes from the conversion of degrees to mas and the conversion from $\sigma$ to FWHM with an extra factor of $\pi$ originating from the Fourier transform. For fitting we use \textsc{emcee} and the same method employed in the geometric modelling of \citet{leftley_parsec-scale_2019}. We model three variables: $\Theta$, $s_f$, and $f$ where $f$ is the fractional amount for which the variance is underestimated by the likelihood function if the errors were assumed correct \citep{foreman-mackey_emcee:_2013}. We do not employ the point source fraction from their modelling since visibility squared shows no clear evidence of a constant. However, a point source component is predicted due to contributions from the accretion disk; we do not see any sign of a turnover due to the point source because the extended component is not sufficiently resolved. To compare to the reverberation mapped radius from Boulderstone et al. (in prep), we utilise the method in \citet{gravity_collaboration_resolved_2020}, in which a fitted Gaussian FWHM is converted to a ring radius assuming a fixed point source contribution. While it is possible to directly fit a ring model, a Gaussian model allows us to keep our models consistent and comparable when searching for more complex geometry in latter modelling. Similarly to \citet{gravity_collaboration_resolved_2020}, when converting from Gaussian FWHM to ring size we do include the expected point source contribution. We use a point source luminosity fraction of 0.15 instead of the 0.2 they report. A point source fraction of 0.15 is chosen here because the point source fraction of 0.2 is a general value determined by \citet{kishimoto_innermost_2007} but a specific value for ESO\,323-G77 was later determined by Kishimoto et al.\ (in prep).

\begin{figure}
    \centering
    \includegraphics[width=0.49\textwidth,trim={0cm 0 0 0},clip]{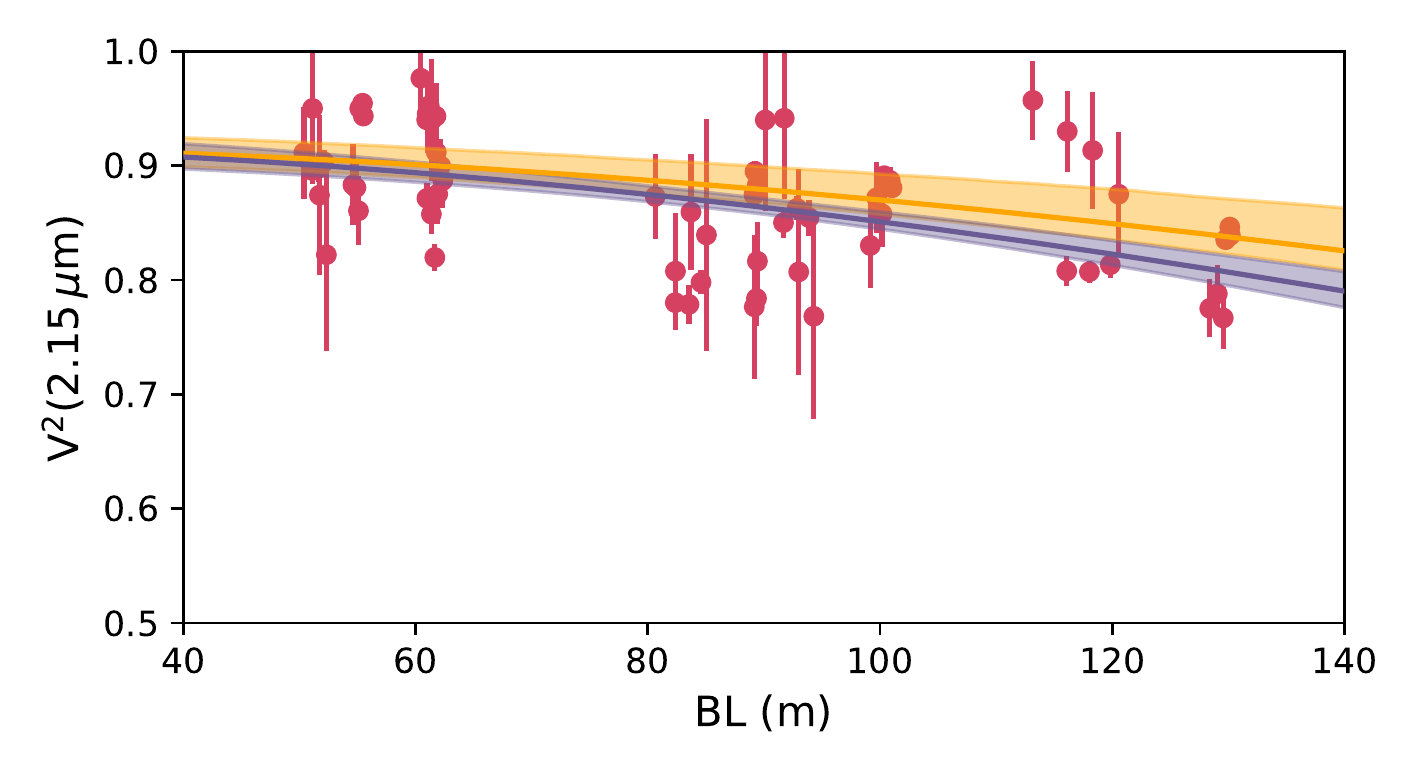}
    \caption{The visibility squared observations, at 2.15$\,\mu$m, are in red. In purple is the best fit model to the observations with 1$\,\sigma$ error and in orange is the reverberation mapped radius with 1$\,\sigma$ error. We only show visibility squared from 0.5 to 1 for clarity.}
    \label{fig:ReverbComp}
\end{figure}

From the fitting of the 2.15$\,\mu$m bin, we find a FWHM of the Gaussian of $0.44^{+0.05}_{-0.05}$\,mas (Figure\,\ref{fig:ReverbComp}). Using an angular size for ESO\,323-G77 of 0.311\,pc\,mas$^{-1}$, to match the distance used for the MIDI modelling in \citet{leftley_new_2018}, this translates to a ring + point source radius of $105^{+11}_{-12}$ light days. The result is within $1\,\sigma$ of the reverberation mapped radius of $89^{+11}_{-18}$\,days which also assumes a thin ring model. The 2.15$\,\mu$m bin was chosen because the central wavelength best matches the reverberation mapping wavelength.

{}

\begin{figure}
    \centering
    \includegraphics[width=0.49\textwidth,trim={0cm 0 0 0},clip]{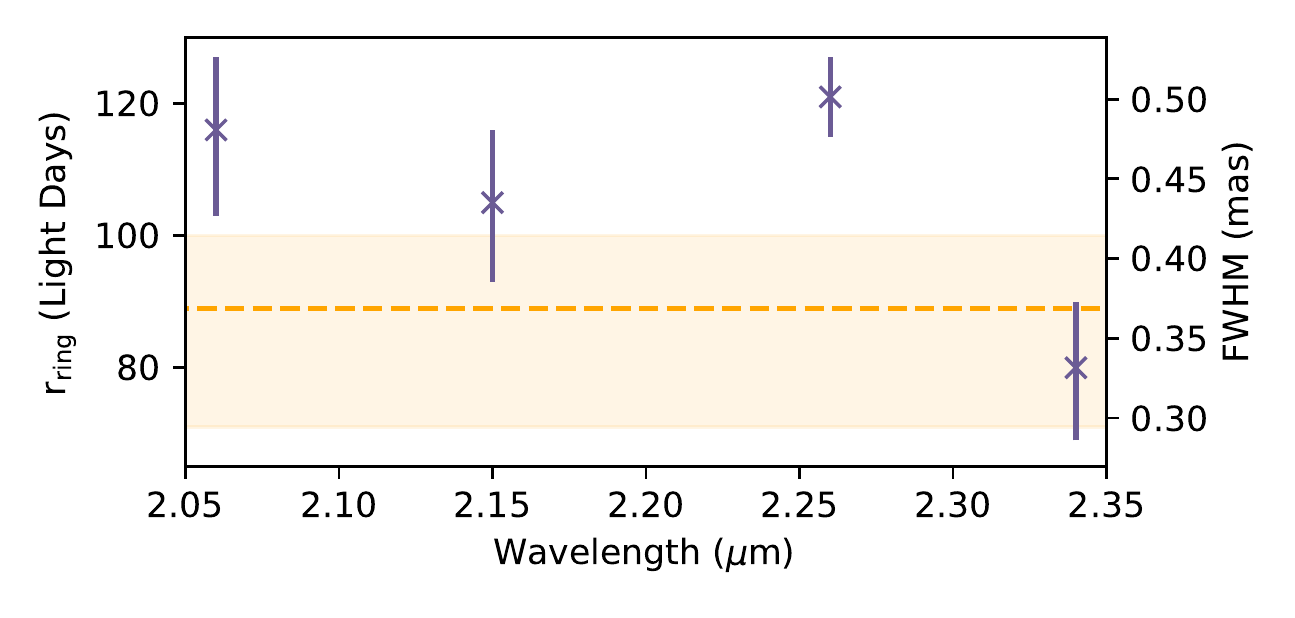}
    \caption{The results of the circular Gaussian fits presented in Table\,\ref{tab:ReverbMap}. r$_\mathrm{ring}$ on the left of the plot is the radius of the thin ring calculated from the fitted Gaussian full width half maximum (FWHM) on the right. The points in purple are the fits to the interferometric data, the orange area is the size measured from reverberation mapping.}
    \label{fig:ReverbMapPlot}
\end{figure}

As a further test, we make the assumption that the regions probed by each wavelength bin in the GRAVITY data are the same within the angular resolution of GRAVITY. We fit all the wavelength bins, except the 1.99$\,\mu$m bin, in the same manner as the 2.15$\,\mu$m bin; the resulting ring radii are given in Table\,\ref{tab:ReverbMap} and shown in Figure\,\ref{fig:ReverbMapPlot}. In addition to the 2.15$\,\mu$m bin, we find that the radial size agrees within 1$\,\sigma$ of the reverberation mapped size for the 2.34$\,\mu$m bin. The remaining two wavelength bins are instead within 2\,$\sigma$ of the reverberation mapped size. Hence, there is a good agreement between both methods; although, the visibility size is slightly larger for three of four bins (see Figure\,\ref{fig:ReverbMapPlot}). A larger interferometric size, as compared to the reverberation mapped size, is expected because interferometry gives the average size of the brightness distribution while reverberation mapping is more sensitive to the inner edge of the hot dust disk which is necessarily smaller \citep{honig_dust-parallax_2014}. If the 2.15$\,\mu$m bin is assumed to be the best candidate for matching the reverberation mapped radii and the disk of ESO\,323-G77 is well described by a thin ring then a distance to ESO\,323-G77 can be calculated by matching the two results. Comparing the sizes results in a distance of $54^{+11}_{-11}$\,Mpc which is within 2\,$\sigma$ of the current luminosity distance of 70$\pm$5\,Mpc although the errors are large. A larger angular radii gives a smaller distance so the 2.06$\,\mu$m and 2.26$\,\mu$m agree less with current cosmology and the best agreement with current cosmology comes from the 2.34\,$\mu$m bin. The calculation is done naively because we do not account for the difference in the region probed by the two techniques. Simultaneous modelling of the light curves used for reverberation mapping, and the interferometric data must be performed to determine the true source geometry and a more accurate distance \citep{honig_dust-parallax_2014}. Furthermore, we ignore the differences between the different distance measures in cosmology and the true geometry of the source which are negligible at such low redshift.

\begin{table}
    \centering
    \begin{tabular}{c | c c}\hline
         Wavelength&Ring Radius& FWHM \\
         ($\,\mu$m)&(Light Days)&(mas)\\\hline \hline\\[-10pt]
         Reverberation&$89^{+11}_{-18}$&$0.37^{+0.05}_{-0.07}$  \\[3pt]
         2.06&$116^{+11}_{-13}$&$0.48^{+0.05}_{-0.05}$\\[3pt]
         2.15&$105^{+11}_{-12}$&$0.44^{+0.05}_{-0.05}$\\[3pt]
         2.26&$121^{+6}_{-6}$&$0.5^{+0.02}_{-0.02}$\\[3pt]
         2.34&$80^{+10}_{-11}$&$0.33^{+0.04}_{-0.05}$\\[3pt]
         Average&$100^{+5}_{-5}$&$0.41^{+0.02}_{-0.02}$
    \end{tabular}
    \caption{The size of the hot dust in ESO\,323-G77 from a thin ring model fit to the visibility squared and the hot dust size from reverberation mapping (Boulderstone et al., in prep). The average is the mean of the visibility sizes.}
    \label{tab:ReverbMap}
\end{table}

\subsection{Angular visibility structure}\label{Section:angvis}

\begin{figure}
    \centering
    \includegraphics[width=0.49\textwidth,trim={0cm 0 0 0},clip]{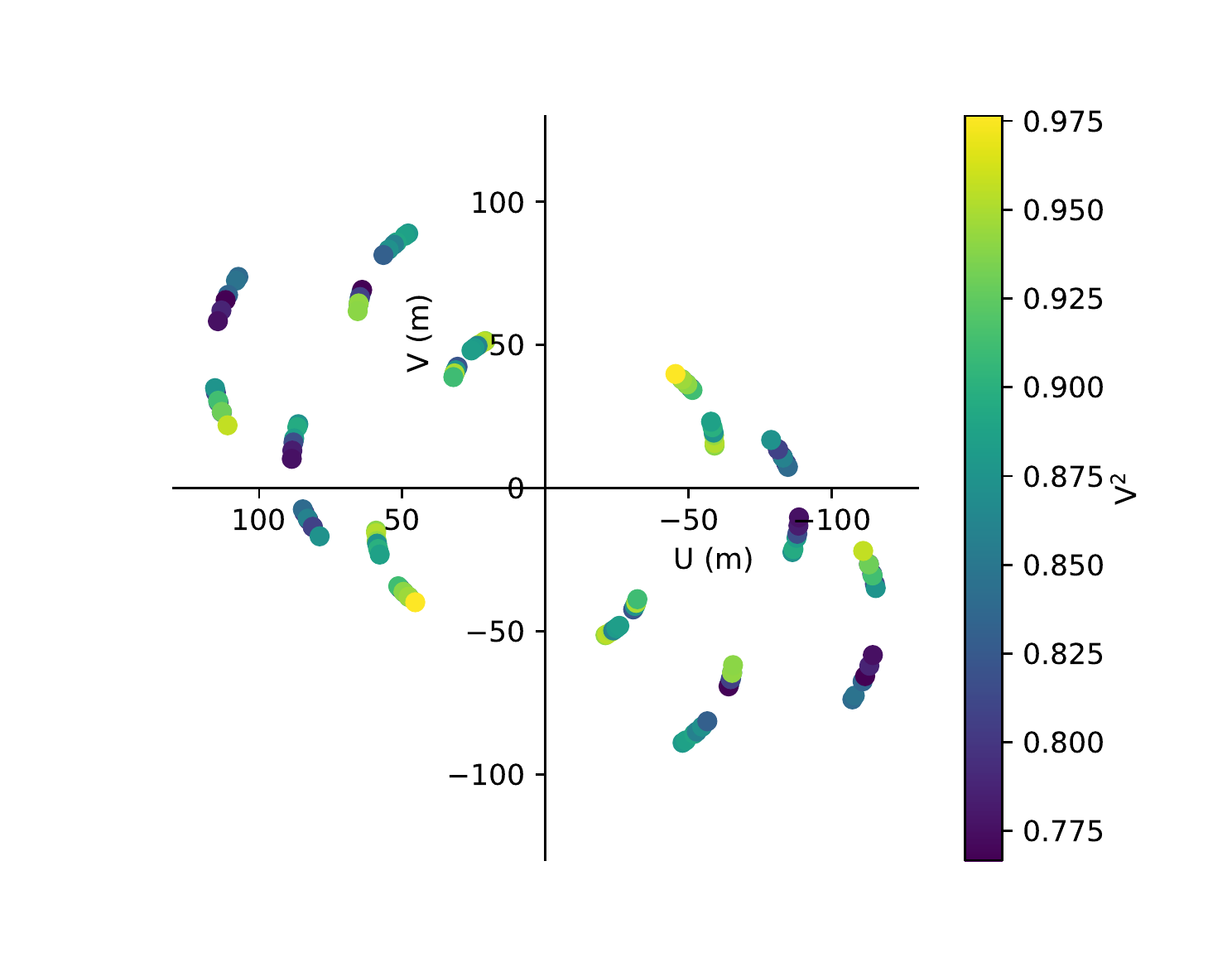}
    \caption{The \textit{uv} plane at 2.15\,$\mu$m, the points are coloured by their squared visibility.}
    \label{fig:scivisUV}
\end{figure}

We plot the \textit{uv} plane for 2.15\,$\mu$m with the points coloured by visibility squared in Figure\,\ref{fig:scivisUV}. From visual inspection, the plot suggests a dependence of visibility on position angle. As a preliminary test for position angle dependency, we attempt to fit the visibility squared data of the 2.06$\,\mu$m, 2.15$\,\mu$m, 2.26$\,\mu$m, and 2.34$\,\mu$m bins individually with a centro-symmetric elongated Gaussian and point source model. The point source is fixed to 0.15 to match the previous modelling and the method follows \citet{leftley_new_2018}. Although the point source is fixed, the introduction of the scale factor means that the number of free parameters is the same as their work. Fitting the latter four wavelength bins individually shows that a non-radially symmetric Gaussian is preferred for all but the 2.06$\,\mu$m bin. The distribution of the fitted parameters for each of the fits can be found in Section\,\ref{sec:AppEllGaussStats} of the Appendix. The 2.06$\,\mu$m bin prefers a radially symmetric Gaussian. The minor to major axis ratio of the elongated Gaussian ($\epsilon$) and the position angle of the major axis (PA) for the other bins can be found in Table\,\ref{tab:El_Gauss_result}.

\begin{table}[]
    \centering
    \begin{tabular}{c|ccc}
         Wavelength ($\mu$m)&$\epsilon$&PA ($^\circ$)&$\Theta_y$ \\ \hline\hline
         2.06*&$1$&$13^{+31}_{-27}$&$0.46^{+0.07}_{-0.11}$\\[3pt]
         2.15&$0.72^{+0.1}_{-0.11}$&$24^{+19}_{-22}$&$0.42^{+0.06}_{-0.09}$\\[3pt]
         2.26&$0.7^{+0.05}_{-0.04}$&$51^{+3}_{-5}$&$0.57^{+0.02}_{-0.02}$\\[3pt]
         2.34&$0.72^{+0.11}_{-0.11}$&$49^{+13}_{-23}$&$0.36^{+0.04}_{-0.06}$\\[3pt]
         $2.15-2.34$&$0.73^{+0.06}_{-0.05}$&$46^{+5}_{-9}$&$0.48^{+0.02}_{-0.02}$
    \end{tabular}
    \caption{The results of the elongated Gaussian and point source fit. $\epsilon$ is the minor to major axis ratio, PA is the position angle of the major axis and $\Theta_y$ is the FWHM of the major axis. *) While the object is not extended within errors a PA is provided for the cases where it is fitted with an extension.}
    \label{tab:El_Gauss_result}
\end{table}

The polar axis of the AGN system, i.e. the axis perpendicular to the plane of the accretion disk, is $174^\circ\pm2^\circ$ or $155^\circ\pm14^\circ$ East of North as defined by polarisation measurements \citep{schmid_spectropolarimetry_2003,smith_seyferts_2004,batcheldor_nicmos_2011} and the mid-IR extension from MIDI \citep{leftley_new_2018}, respectively. Therefore, the dust disk should have a position angle of $84^\circ\pm2^\circ$ or $65^\circ\pm14^\circ$ assuming there is no misalignment between the dust disk and the respective responsible medium. When an object is equatorially scattered, the accretion disk is the scattering medium giving a very good descriptor of the accretion disk's equatorial position angle \citep{smith_seyferts_2004}. However, ESO\,323-G77 is polar scattered, the scattering medium in this case is thought to be dust clouds along the polar axis \citep{schmid_spectropolarimetry_2003}. In the polar scattered case, the determined axis may not be a good descriptor of the precise polar axis. The MIDI extension traces the warm ($\sim$300\,K) dust, the location of which is model dependent. In previous work, it has been found that the warm dust is generally polar extended, although often somewhat misaligned with the polarisation determined polar axis \citep{lopez-gonzaga_mid-infrared_2016}.

The results of the fitting show that the latter two wavelength bins prefer an extension that is closer to equatorial than polar, for both polar axis measurements, whilst the 2.15$\,\mu$m bin is closer to polar by polarisation and equatorial by MIDI; although, the 2.15$\,\mu$m bin has relatively large uncertainty and is within 3\,$\sigma$ of both directions. While the position angle of the 2.26$\,\mu$m and 2.34$\,\mu$m bins agree within 1$\,\sigma$ of the MIDI equatorial direction, none of the results agree within $1\,\sigma$ uncertainty of the equatorial direction of $84^\circ\pm2^\circ$ or the corresponding polar direction.

\begin{figure}
    \centering
    \includegraphics[width=0.45\textwidth,trim={0cm 1cm 0cm 2cm}, clip]{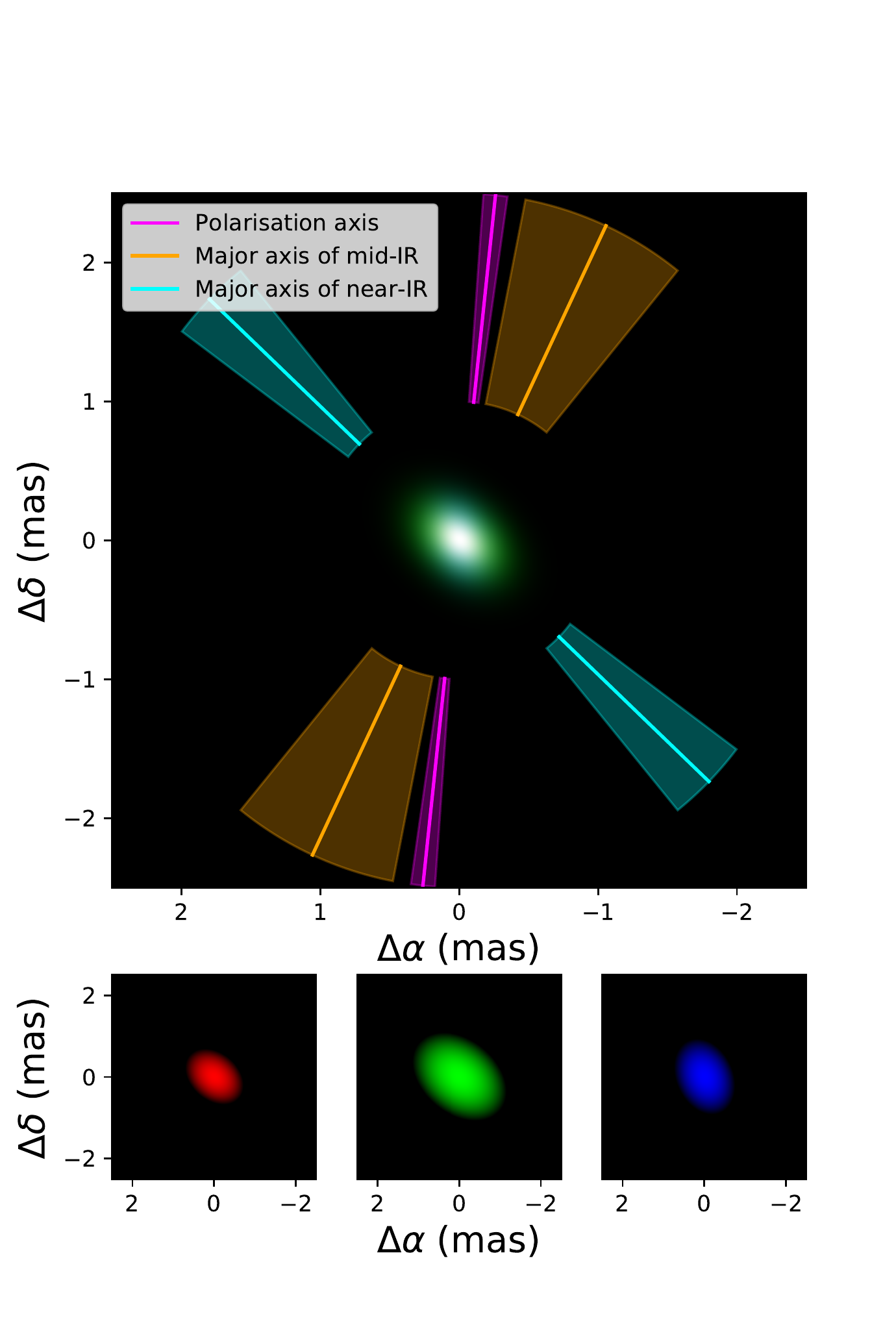}
    \caption{The overlapped elongated Gaussian fits to the squared visibility at 2.15$\,\mu$m, 2.26$\,\mu$m, and 2.34$\,\mu$m compared to the mid-IR extension direction and the polar axis from polarisation.  The 2.15$\,\mu$m, 2.26$\,\mu$m, and 2.34$\,\mu$m wavelength bins are plotted in blue, green, and red respectively.}
    \label{fig:RGB}
\end{figure}

In an attempt to reduce the uncertainties, we perform a multi-wavelength fit. This fit is performed using the 2.15$\,\mu$m, 2.26$\,\mu$m, and 2.34$\,\mu$m bins under the assumption that the different wavelength bins probe the same dust with different angular resolution. The result of this fit is an elongated Gaussian with an axis ratio of $0.73^{+0.06}_{-0.05}$ and a position angle of $46^{+5}_{-9}$. The result is inconsistent with both the polarisation determined polar axis and equatorial axis (see Figure\,\ref{fig:RGB}); however, it is closer to equatorial than polar. The position angle is consistent with the equatorial direction determined with MIDI.

{}

\begin{figure*}[t]
    \centering
    \includegraphics[width=0.9\textwidth]{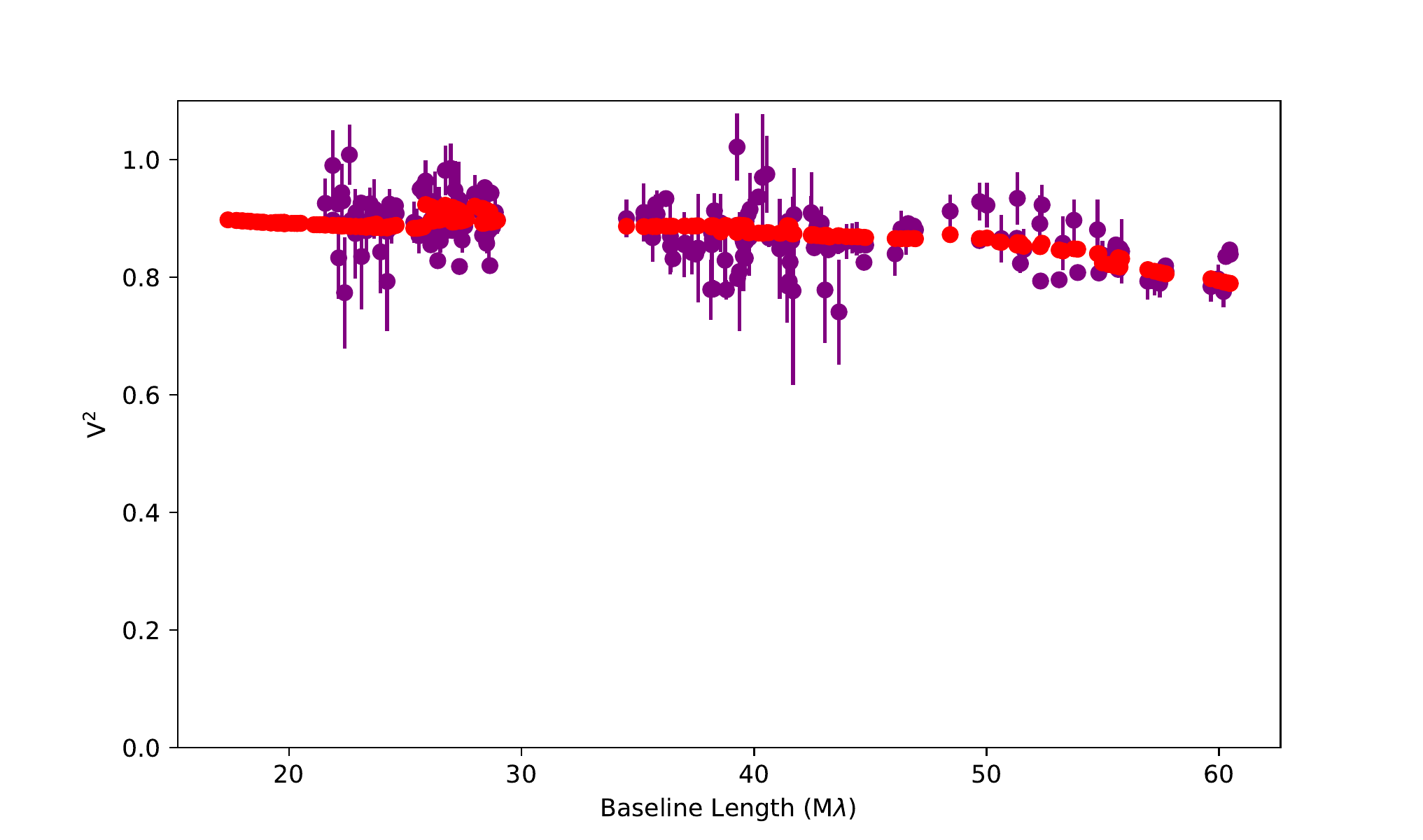}
    \caption{The visibility squared for the 2.15$\,\mu$m, 2.26$\,\mu$m, and 2.34$\,\mu$m wavelength bins are plotted in purple against triangle perimeter in M$\lambda$. The red depicts the best fitting model at sampled at the same \textit{uv} locations.}
    \label{fig:VisMod}
\end{figure*}{}

\section{Modelling}\label{sec:Grav_Mod}

Simple models fit to the squared visibility gave us an overview of the dominant source structure; we will now proceed to model the visibility and closure phase simultaneously. Fitting visibility and phase will allow us to extract more information from the observations. The modelling in this chapter was performed in \textsc{python} and makes use of the packages: \textsc{astropy} \citep{astropy_collaboration_astropy:_2013}, \textsc{galario} \citep{tazzari_galario_2018}, and\textsc{ emcee}. \textsc{galario} allows us to perform Fourier transforms on images and derive complex visibility for a given set of \textit{uv} points with little computational expense. Less computational expense makes expensive fitting methods such as MCMC compatible with highly flexible models. For the MCMC method we use \textsc{emcee} with flat priors for all given variables.

We attempt to explain the GRAVITY observations in the three longest wavelength bins simultaneously. Any attempt to fit the wavelength bins separately failed to constrain any reliable geometry. In a single wavelength bin, the data are too sparse for the relatively complex models, as compared to the ones used in Section\,\ref{Section:ObRes}. Therefore, we make the assumption that the narrow range of wavelengths covered by the GRAVITY fringe tracker probes the same hot dust component at different baseline lengths, essentially allowing us to fill more of the \textit{uv} plane. For simplicity, the models in this section are fitted in wavelength space.

We attempt to recreate the squared visibility and closure phase simultaneously using point sources with a method analogous to clean-like image reconstruction. To compare different numbers of point sources, the Bayesian Information Criteria (BIC) is used. The BIC is defined as:

\begin{equation}
    \begin{array}{GGG}
         \mathrm{BIC}&=&\ln{(n)}\,k-2\,\ln{(L)},
    \end{array}{}
\end{equation}{}

where $n$ is the number of utilised data points, $k$ is the number of parameters of the model, and $L$ is the likelihood: 

\begin{equation}
    \begin{array}{GGG}
         2\ln{(L)}&=&-\sum_m^N\left[\frac{(y(x_m)-\mathrm{model}(x_m,\alpha))^2}{\sigma_m^2+f^2\mathrm{model}(x_m,\alpha)^2}+\right.\\
         &&\left.\ln\left(2\pi\left(\sigma_m^2+f^2\mathrm{model}(x_m,\alpha)^2\right)\right)\vphantom{\frac{y^2}{y^2}}\right],
    \end{array}{}
\end{equation}{}
for a set of data of length N with values $y$, positions $x$ and a model with parameters $\alpha$. The variance is denoted by $\sigma$ and $f$ is the fractional amount for which the variance is underestimated by the likelihood function if the errors were assumed correct \citep{foreman-mackey_emcee:_2013}.

Each point source has a $\Delta$RA, $\Delta$Dec, and point source fraction ($p_f$). A point source in this case is not a delta function but a 2D Gaussian with a FWHM of a pixel which prevents an image with finite resolution producing a pixelated probability space. After the squared visibility is calculated, a scale factor ($s_f$) is introduced in the same manner as Equation\,\ref{eq:GaussAny}. The sum of all point source fractions is defined to be one. A single point source was fixed to the central pixel with a fixed fraction of 0.15, the value used for the accretion disk for the reverberation mapping comparison. All position parameters are made relative to the central point source. The point source model contained $k = 3(\mathrm{N}_p-1)$ free parameters, not including $s_f$, where $\mathrm{N}_p$ is the number of points. When calculating the BIC the number of free parameters does not include $s_f$ because we are interested in the relative BIC, not the absolute, and it is a non-physical component that is present in all models and only effects the visibility squared. The bounds for the free parameters are given in Table\,\ref{tab:point_bounds}. When creating an image from the point source model, the model is shifted so the central image pixel is the location of the mean of all point source positions weighted by point source fraction to reduce boundary effects when performing the Fourier transform due to the finite image size. The bounds set for each parameter were as follows:

\begin{table}[h]
    \centering
    \begin{tabular}{l|cc}
        \hline
         Parameter&Lower&Upper  \\ \hline\hline
         $\Delta$RA&-10\,mas&10\,mas\\
         $\Delta$Dec&-10\,mas&10\,mas\\
         $p_f$&0&0.85
    \end{tabular}
    \caption{The bounds for each free parameter of the point source model.}
    \label{tab:point_bounds}
\end{table}{}

The model sampled from 2 to 32 point sources, the initial walker positions were selected from a uniformly random distribution over the available parameter space. To prevent point source stacking, the bound is set that two point sources may not occupy the same pixel. When using the BIC, two point sources in the same pixel will always be a worse fit than one point source with the summed point source fractions of both due to the use of additional unnecessary parameters. Therefore, it is inefficient to consider this solution. The pixel size is 0.0697\,mas with an image size of 512x512 pixels. The squared visibility and closure phase are then calculated from the image and compared to the data.

The MCMC analysis of each $\mathrm{N}_p$ is run for 4000 iterations with 200 walkers. For the last 2000 iterations, or $4\times10^5$ samples, we calculate the BIC. All BICs from each $\mathrm{N}_p$ are compared and the best combination of values are recorded. The lowest BIC parameter set are then used as the starting point for a final MCMC analysis by using 200 walkers scattered by a normal distribution of $\sigma=10^{-2}$ centred on the lowest BIC parameter set for 4000 iterations. The last 2000 iterations are used to calculate the uncertainties and position of each parameter. We use the described method because the relatively sparse data coverage, as compared to the GRAVITY imaged AGN NGC\,1068 \citep{gravity_collaboration_image_2020}, requires an extremely large number of iterations to find the true result from a random starting position. There are also many "local minima" in the probability space preventing us from finding a good starting position with a minimiser. Furthermore, it is too computationally expensive to calculate the BIC for every parameter combination. Therefore, the given method provided the best trade-off between accuracy and computational expense for the data.

Utilising the same fitting method, we also employ a second set of more complex models. The second set of models are composed of Gaussians with variable FWHMs instead of the fixed FWHM point sources. Both elongated and radially symmetric Gaussians were attempted. These models are described in more detail in Section\,\ref{sec:Appendix_models} in the Appendix. Both Gaussian models were compared to the point source model using the BIC and provide no improvement with the current available data.

\section{Modelling Results and Discussion}\label{Sec:ModRes}

\begin{figure}
    \centering
    \includegraphics[width=0.5\textwidth]{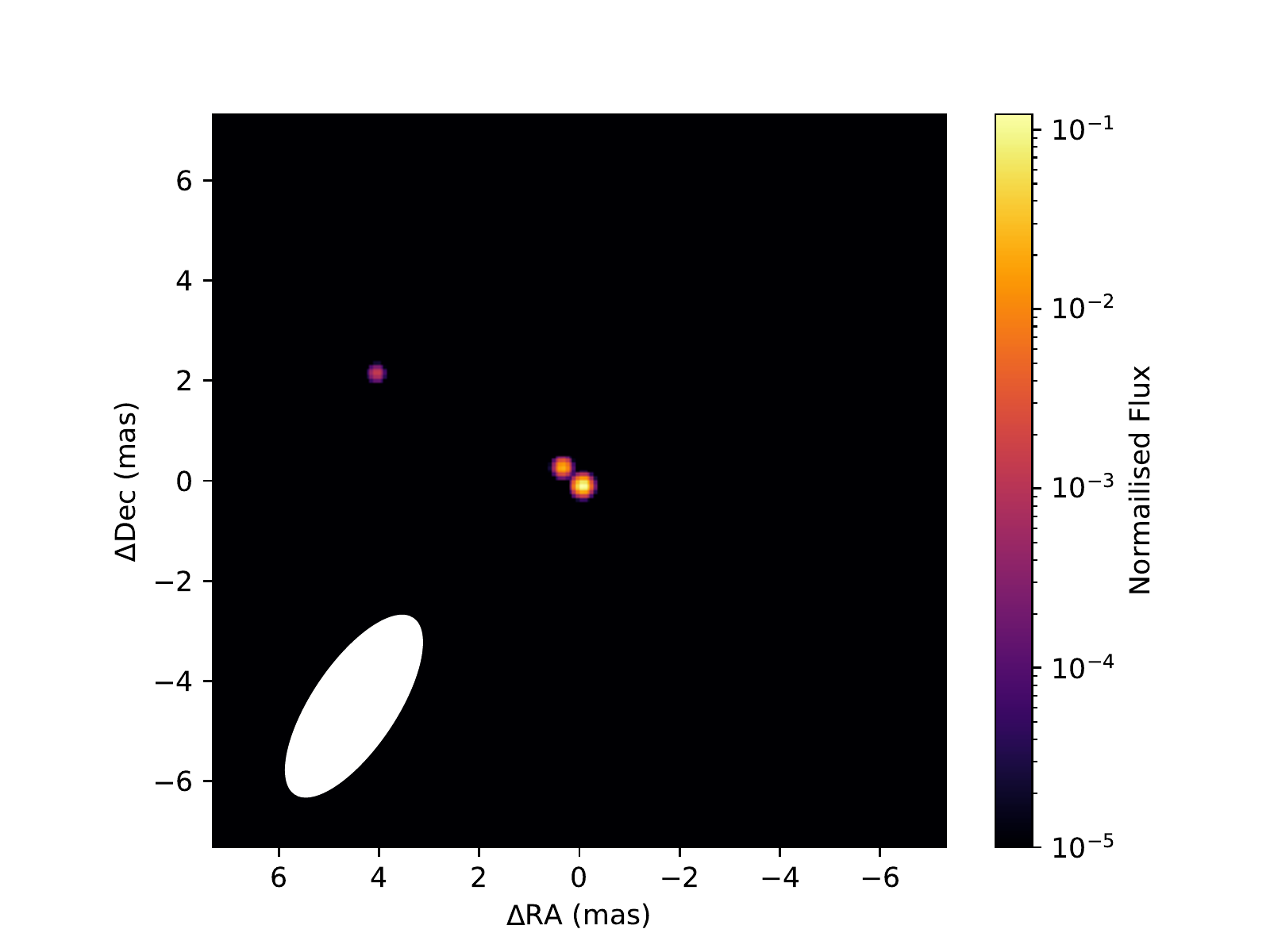}
    \caption{The three point source best fit model to the GRAVITY observations. This image is not convolved with the beam size. The colourbar represents normalised flux in log scale. The beam is given in white.}
    \label{fig:grav_mod_pic}
\end{figure}

We find that the best fitting model, out of the three different models presented, is the point source model using 3 point source components (Table\,\ref{tab:GravModRes}). The point source model provides a good description of the closure phase (Figure\,\ref{fig:GravClosMod}) and the squared visibility (Figure\,\ref{fig:VisMod}). The model, unconvolved with the beam size, is shown in Figure\,\ref{fig:grav_mod_pic}. We find that two of the sources, one of which is the fixed component, are separated by $0.5\,$mas and responsible for $99\%$ of the total flux. While the separation is less than the Rayleigh criterion for resolution of $\frac{\lambda}{2\mathrm{B}_\mathrm{max}}$ where $\lambda$ is the wavelength and $\mathrm{B}_\mathrm{max}$ is the maximum baseline length, this resolution denotes the separation at which the correlated flux of two point sources becomes zero. In reality, the resolution depends on the signal to noise of the observations and partially resolved structure can be constrained through modelling on scales less than the beam size. From the Gaussian modelling of the visibility squared, we determine that the physical size at the ~0.5\,mas scale can be robustly constrained. Therefore, we conclude the measured separation is reliable. The central pair lie along a position angle of $50^\circ$ which agrees more with an equatorial extension than a polar extension. This would suggest that there is structure on the $0.5\,$mas scale in an equatorial direction. The direction agrees with that found with the elongated Gaussian fits and the separation agrees with the size from reverberation mapping. In previous work it has been shown that an elongated Gaussian, for the \textit{uv} plane of ESO323-G77, can be constrained in any direction with the fitting method utilised \citep{leftley_new_2018}. Therefore, we conclude that the direction is reliable and the structure could correspond to the accretion disk and the dust disk or to either side of the dust disk's inner rim.

The third component is $4.2\,$mas away from the fixed component and is responsible for 1\% of the flux. The faint component is essential for explaining the closure phase. It causes the small shift in photocentre required at lower spatial resolutions. It resides to the North-East of the group which is, temptingly, $\sim90^\circ$ from the mid-IR extension found in \citet{leftley_new_2018}. However, it is on a larger scale than expected of the dusty disk and ESO\,323-G77 is thought to be inclined at $60^\circ$ from face on, when modelled with CAT3D-WIND \citep{honig_dusty_2017,leftley_new_2018}, which makes interpretation more difficult. Therefore, it would be misleading to claim that the faint component is in anyway related to the dusty disk at this stage. Alternatively, the blob may not be true geometry but instead it could be indicative of an asymmetry or substructure in the equatorial direction. In a three point source model, the distance to the faint point is well constrained; however, this does not rule out a more complex asymmetry on a different spatial scale. For example, a non uniform distribution of dust clumps in the disk or a warm dust blob further from the sublimation radius. This would require further modelling to check. 

{}

\subsection{Comparison of results to model predictions}

In terms of spatial distance, the central point sources are separated by $\sim0.17\,$pc using 0.311\,pc\,mas$^{-1}$. The separation distance is remarkably close to the expected K band size of the hot dust from the \textit{CAT3D-WIND} model \citep[][]{honig_dusty_2017} given in \citet{leftley_new_2018} of 0.15\,pc. The  size is predominantly defined by the drop of the squared visibility. The central separation is essentially what was measured when we calculated the ring size from the visibility squared in Section\,\ref{Sec:Reverb}. With the limited \textit{uv} coverage the two point sources would appear similar to a ring.

\begin{figure}
    \centering
    \includegraphics[width=0.5\textwidth,trim={0cm 0 0 0},clip]{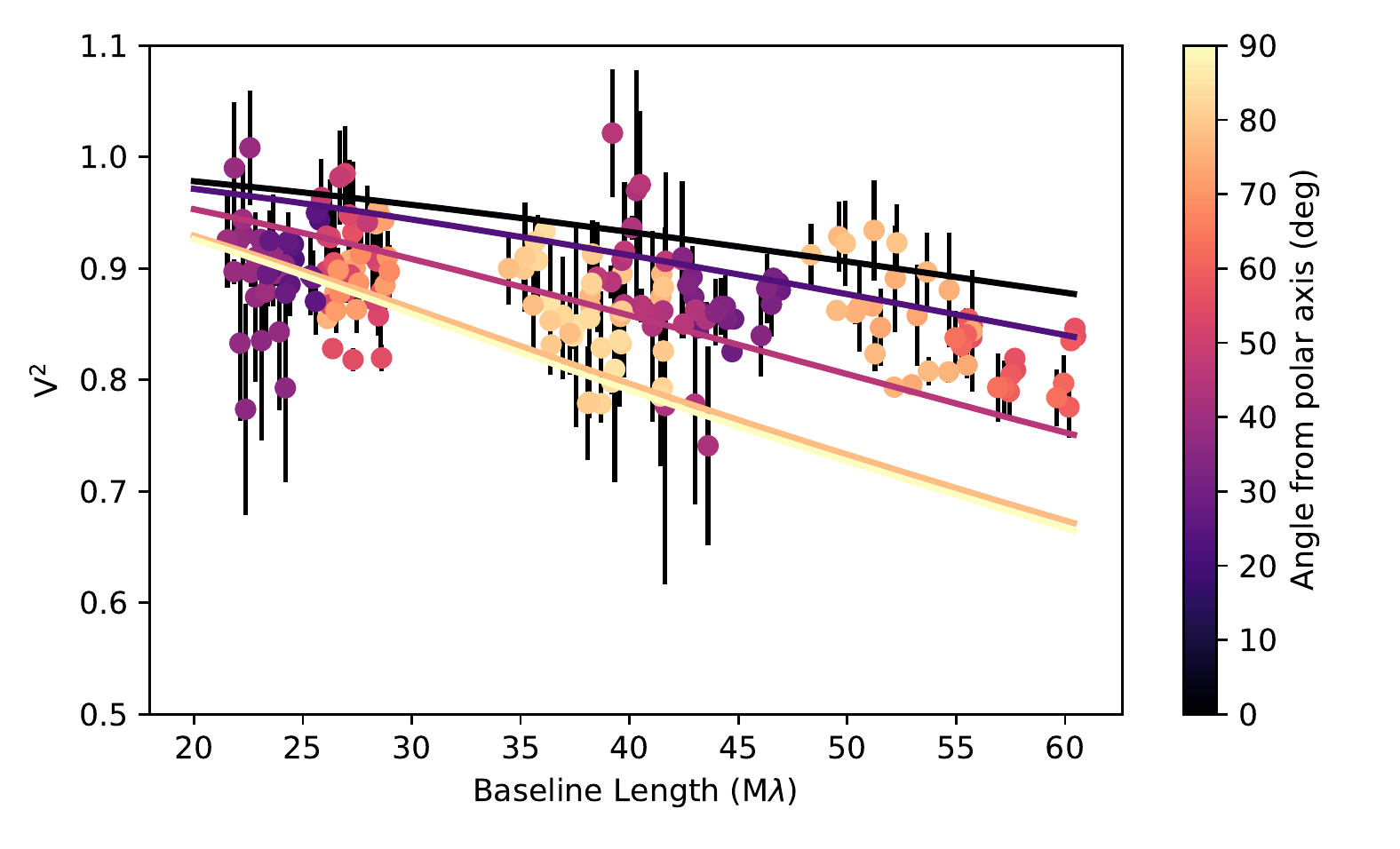}
    \caption{The 2.2\,$\mu$m prediction of the visibility squared made by the \textit{CAT3D-WIND} model created in \citet{leftley_new_2018} compared to the data. The model has a sublimation radius of 0.065\,pc used in their work.}
    \label{fig:cat3d-near-IR}
\end{figure}{}

\begin{table}[hb]
    \centering
    \begin{tabular}{c|ccc}
         Component&p$_\mathrm{f}$&$\Delta$RA&$\Delta$Dec  \\ \hline\hline
         $1*$&0.15&0&0 \\[5pt]
         2&$0.84^{+0.001}_{-0.001}$&$-0.4^{+0.03}_{-0.04}$&$-0.36^{+0.07}_{-0.06}$\\[5pt]
         3&$0.01^{+0.001}_{-0.001}$&$3.7^{+0.06}_{-0.07}$&$1.88^{+0.07}_{-0.08}$
         
    \end{tabular}
    \caption{The modelling result for the combination of the 2.15$\,\mu$m, 2.26$\,\mu$m, and 2.34$\,\mu$m wavelength bins. *This component was fixed.}
    \label{tab:GravModRes}
\end{table}{}

\begin{figure}
    \centering
    \includegraphics[width=0.5\textwidth,trim={0cm 0 0 0},clip]{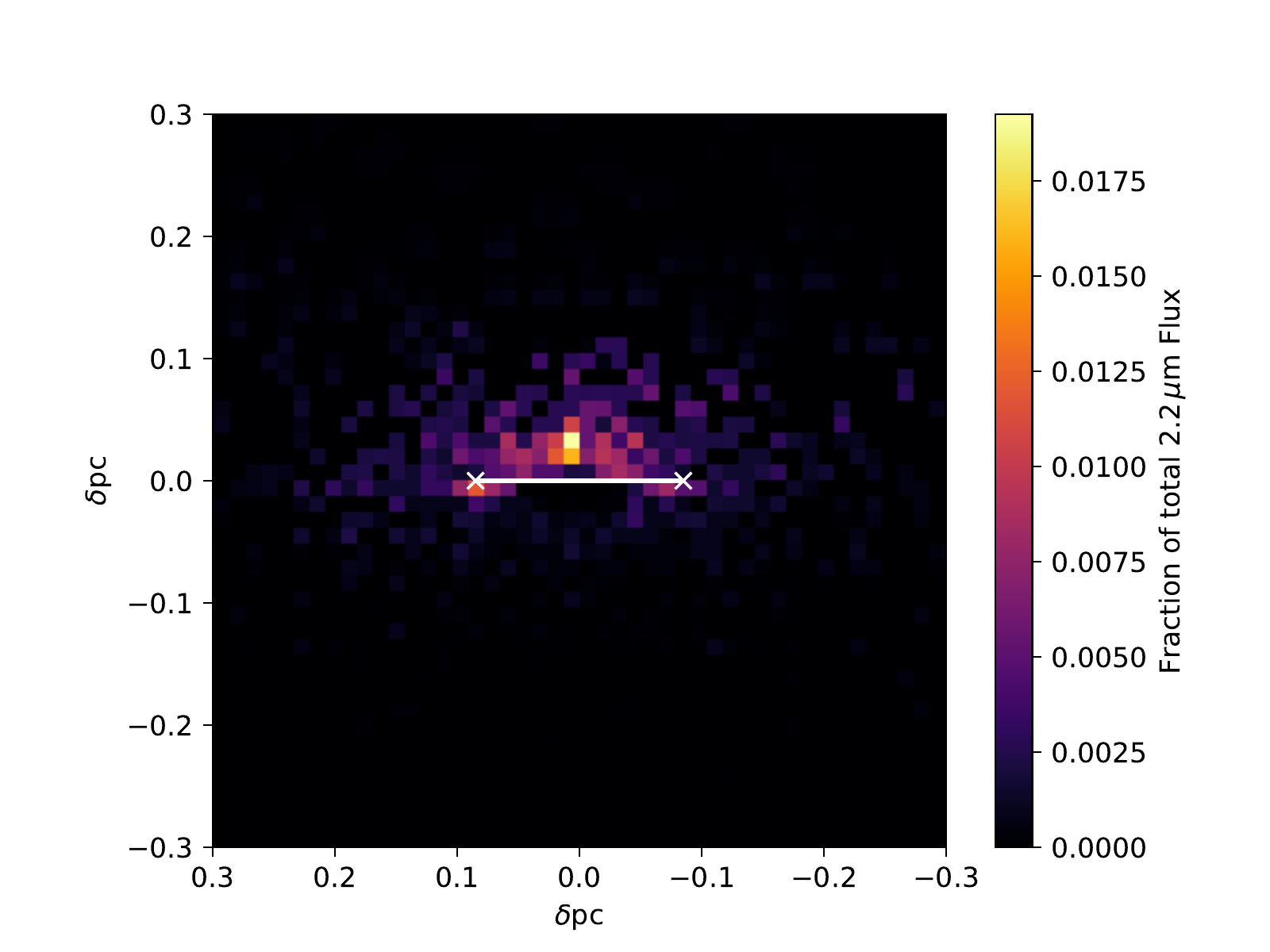}
    \caption{The 2.2\,$\mu$m \textit{CAT3D-WIND} model created in \citet{leftley_new_2018} compared to the central point source separation of the point source model (white). The model has a sublimation radius of 0.065\,pc used in their work.}
    \label{fig:cat3d-im}
\end{figure}

The \textit{CAT3D-WIND} model that was created in \citet{leftley_new_2018} can be used to create what would be seen by a near-IR interferometer in the same manner as was performed for the mid-IR. The \textit{CAT3D-WIND} prediction is shown in Figure\,\ref{fig:cat3d-near-IR}. The \textit{CAT3D-WIND} model is consistent with the data although an improvement to the fit can be found by scaling the model size from the predicted sublimation radius of 0.065\,pc to 0.04\,pc. The similarity between the predicted radius of 0.065\,pc and half the point separation in the recovered point source model supports the idea each point is either side of the hot dust disk's inner rim (see Figure\,\ref{fig:cat3d-im}) but does not exclude the accretion disk and hot dust suggestion. While the point source separation is larger than the sublimation radius of the \textit{CAT3D-WIND} model, it expected to be larger because we are using a point source model to explain extended structure and will therefore recover the average brightness size not the innermost rim separation. The central group does also contribute to the closure phase, however, the dominant component is the interplay between the group and the distant faint source.

\subsection{Gaussian models}

The multiple radially symmetric Gaussians model does reproduce the data, however, it is not a better fit than the point source model with the same number of free parameters or fewer. It also does not provide a significantly better fit, as described by the BIC, using a greater number of free parameters. Therefore, the BIC is larger for the radially symmetric Gaussian model when compared to the best point source model in all cases. In the elongated Gaussian model, the axis ratio for every component is unconstrained. Therefore, the point source model is the best descriptor based on the currently available data.

{}

We conclude that the central group of points represents structure on the scale of the putative sublimation radius and hot dust size predicted by the \textit{CAT3D-WIND} in \citet{leftley_new_2018}. The central extension is $73^\circ$ away from the MIDI determined polar axis and $54^\circ$ away from the polarisation axis which suggests the extension is equatorial. This agrees with the elongation found from Section\,\ref{Section:angvis}. If the far point, at $\sim1.3\,$pc from the fixed component, is real structure it could be a hot cloud blown out from the sublimation radius, a large cooler clump, or a more distant hot spot heated by some other means. In the case of the cooler clump, the faint component should be prominent in the L or M band with MATISSE. It may also be indicative of more complex substructure in the disk not captured by the simple models. The data are currently too noisy and sparse to determine if the classical torus model or the dusty wind model is a better explanation of the observations. However, our results support the idea that the K band hot dust is located in an equatorial dust structure consistent with the sublimation radius.

\section{Summary}\label{Sec:Summary}

We present a study of ESO\,323-G77 in the near-IR using the VLTI instrument GRAVITY. We constrain the geometry of the hot ($\sim1000$\,K) dust through the use of simple squared visibility models and more complex squared visibility and closure phase models. Our main conclusions are:

\begin{enumerate}
    \item The average radial size of the hot dust emission, as inferred by the visibility squared, is consistent with the size of the sublimation radius as inferred from K band reverberation mapping.
    
    \item The squared visibility shows angular structure at 2.15\,$\mu$m, 2.26\,$\mu$m, and 2.34\,$\mu$m. The 2.26\,$\mu$m and 2.34\,$\mu$m extensions are more consistent with the equatorial direction than polar when compared to both the polar axis inferred by polarisation and the MIDI extension from \citet{leftley_new_2018}. The extension at 2.15\,$\mu$m is consistent with the equatorial direction as inferred by the MIDI extension but not  polarisation.

    \item The squared visibility and closure phase is best explained by three point sources, when using the BIC test, which are consistent with the interferometric size predicted by the CAT3D-WIND model from \citet{leftley_new_2018,honig_dusty_2017}, the sublimation radius size from reverberation mapping, and the expected direction of the disk.
\end{enumerate}

\acknowledgements
We would like to thank the referee for the kind comments and suggestions which helped improve this work.

JHL, SFH, and DJW acknowledge support from the Horizon 2020 ERC Starting Grant \textit{DUST-IN-THE-WIND} (ERC-2015-StG-677117). JHL acknowledges the support of the French government through the UCA\,JEDI investment in the Future project managed by the National Research Agency (ANR) under the reference number ANR-15-IDEX-01. MK acknowledges support from JSPS under grant 16H05731. PG acknowledges support from STFC and a UGC/UKIERI Thematic Partnership. DA acknowledges funding through the European Union’s Horizon 2020 and Innovation programme under the Marie Sklodowska-Curie grant agreement no. 793499 (DUSTDEVILS).

Based on European Southern Observatory (ESO) observing programme 0103.B-0096(A).

\software{Astropy \citep{astropy_collaboration_astropy:_2013}, Corner \citep{foreman-mackey_corner.py:_2016}, Emcee \citep{foreman-mackey_emcee:_2013}, Galario \citep{tazzari_galario_2018}, Matplotlib \citep{hunter_matplotlib_2007}, Numpy \citep{harris_array_2020}}

\bibliographystyle{apj.bst}
\bibliography{Zotero.bib}

\appendix
\section{Elongated Gaussian modelling statistics plots}\label{sec:AppEllGaussStats}

\begin{figure}[b]
    \centering
    \includegraphics[width=\textwidth,trim={0cm 0 0 0},clip]{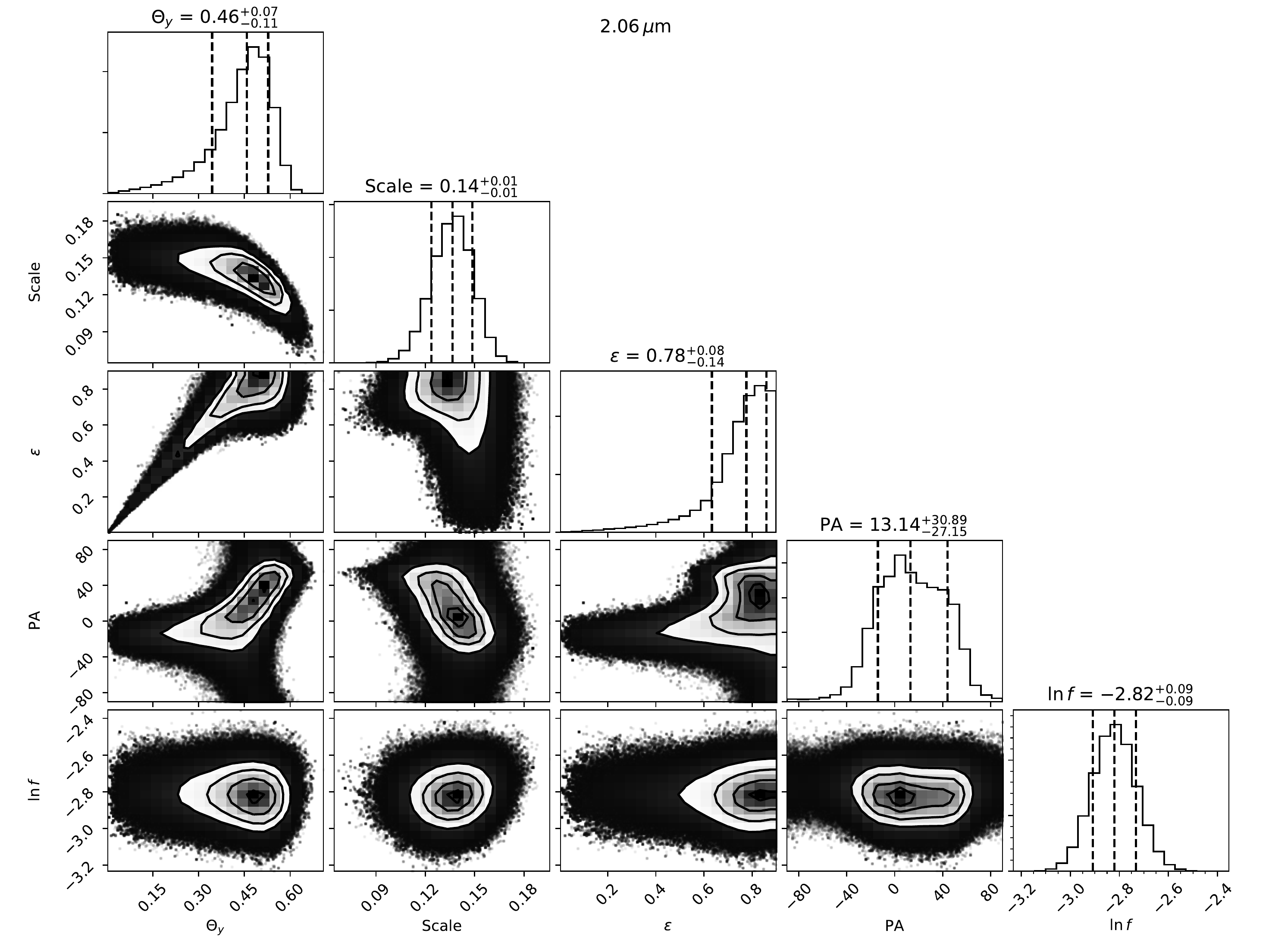}
    \caption{Corner plot \citep{foreman-mackey_corner.py:_2016} of the Probability Density Function (PDF) of each parameter used in the elongated Gaussian fit in Section\,\ref{Section:angvis} for 2.06\,$\mu$m.}
    \label{fig:AppGauss206}
\end{figure}

\begin{figure}[b]
    \centering
    \includegraphics[width=\textwidth,trim={0cm 0 0 0},clip]{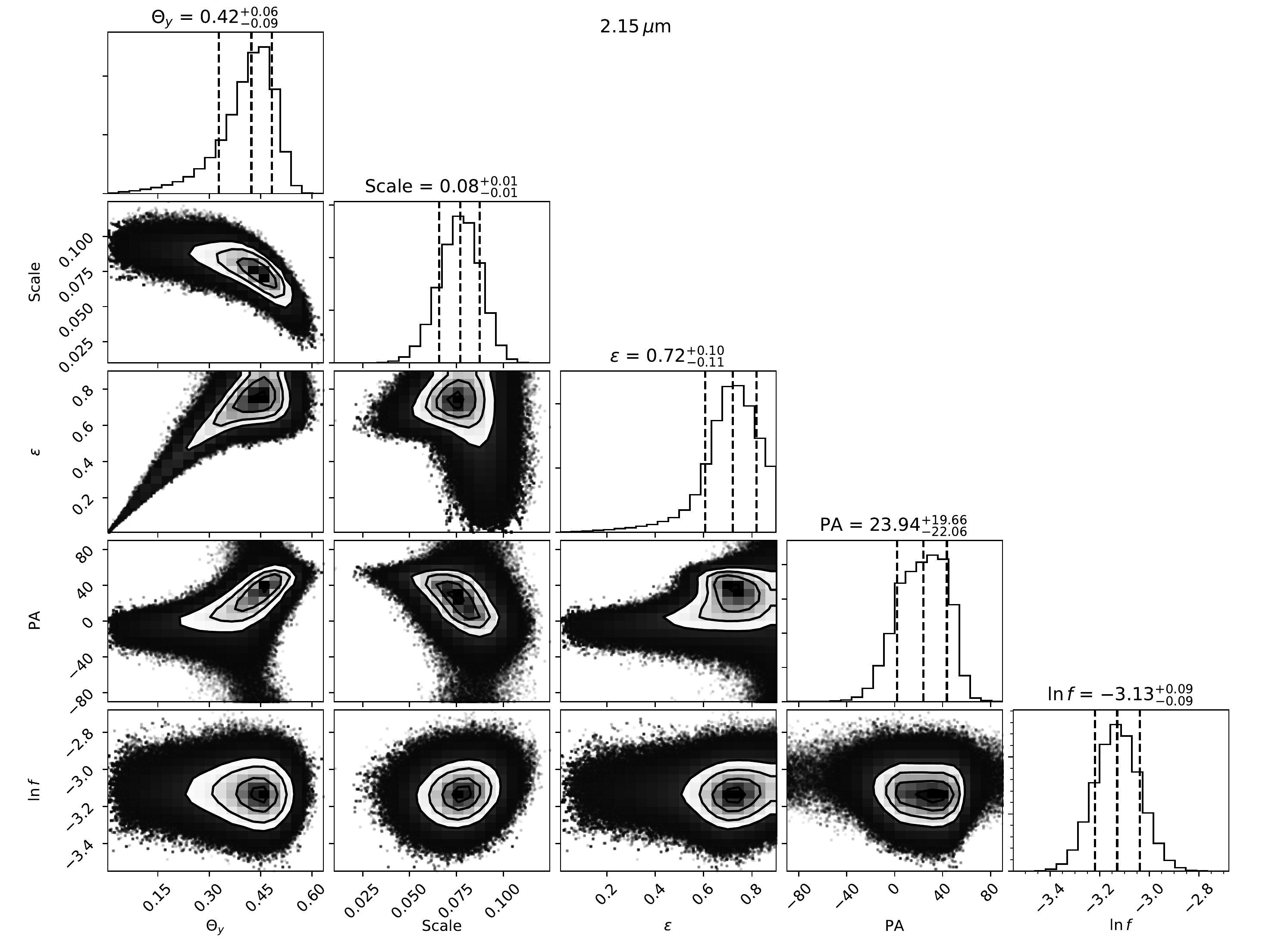}
    \caption{Corner plot \citep{foreman-mackey_corner.py:_2016} of the PDF of each parameter used in the elongated Gaussian fit in Section\,\ref{Section:angvis} for 2.15\,$\mu$m.}
    \label{fig:AppGauss215}
\end{figure}

\begin{figure}[b]
    \centering
    \includegraphics[width=\textwidth,trim={0cm 0 0 0},clip]{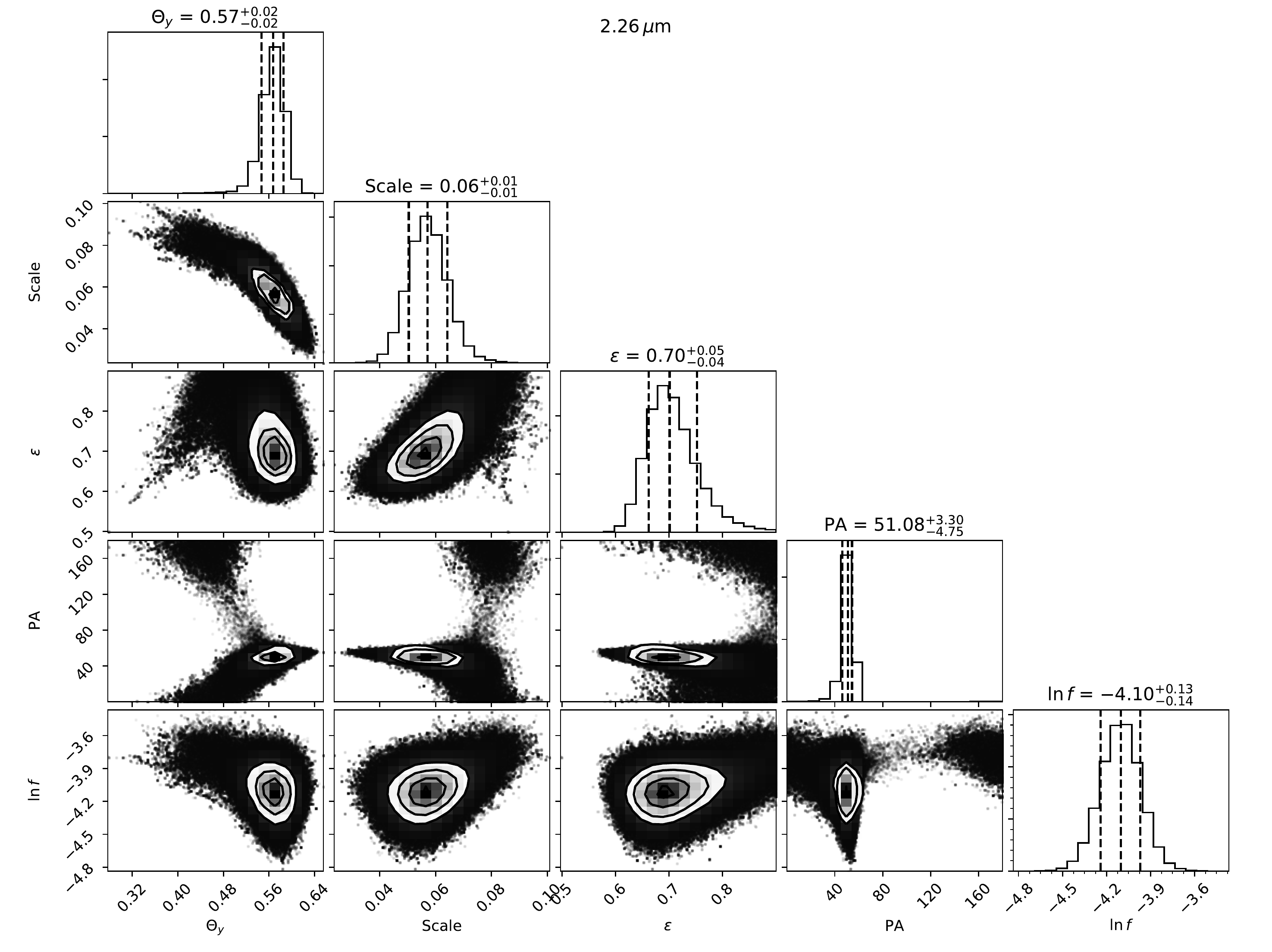}
    \caption{Corner plot \citep{foreman-mackey_corner.py:_2016} of the PDF of each parameter used in the elongated Gaussian fit in Section\,\ref{Section:angvis} for 2.26\,$\mu$m.}
    \label{fig:AppGauss226}
\end{figure}

\begin{figure}[b]
    \centering
    \includegraphics[width=\textwidth,trim={0cm 0 0 0},clip]{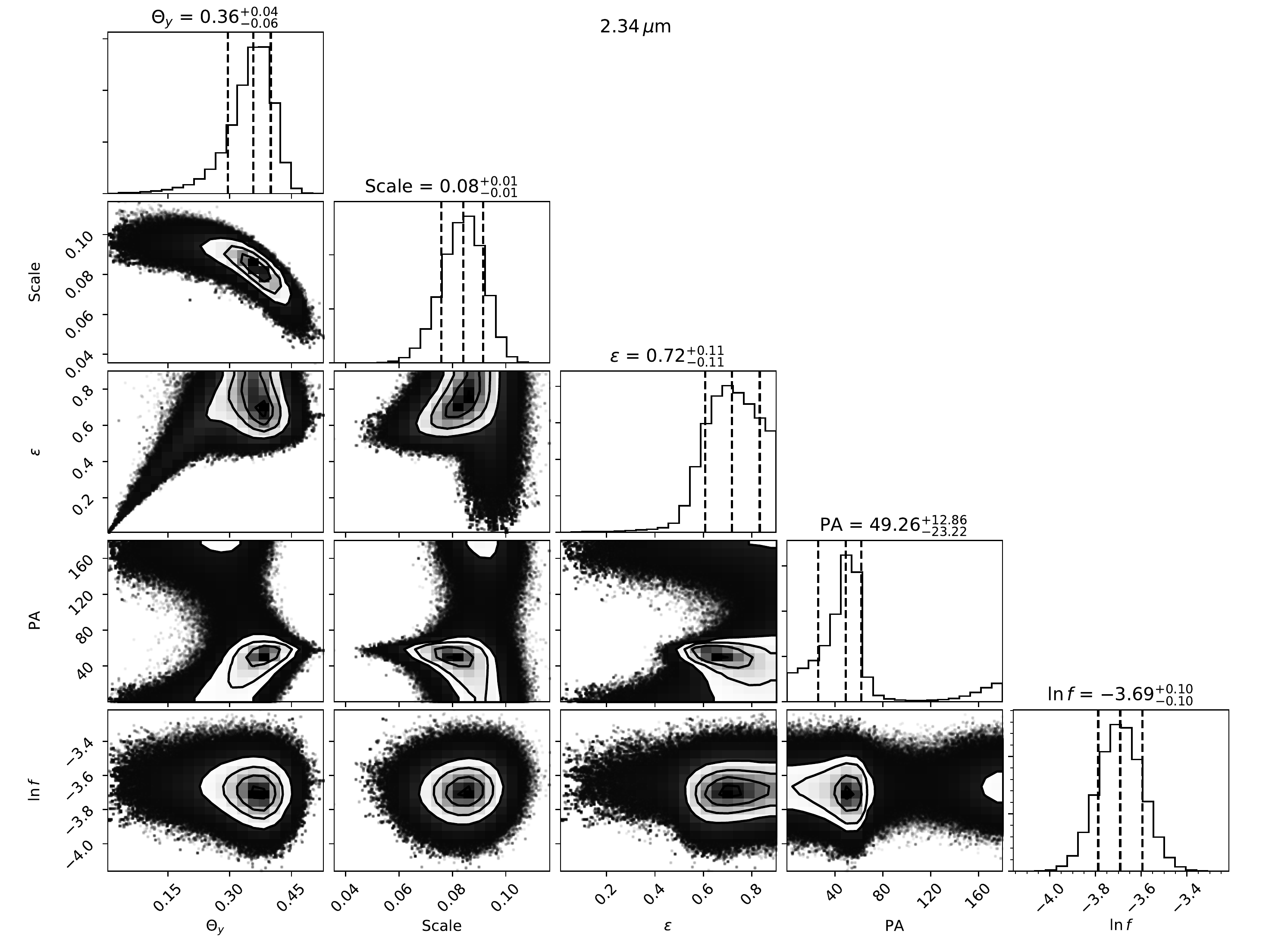}
    \caption{Corner plot \citep{foreman-mackey_corner.py:_2016} of the PDF of each parameter used in the elongated Gaussian fit in Section\,\ref{Section:angvis} for 2.34\,$\mu$m.}
    \label{fig:AppGauss234}
\end{figure}

\begin{figure}[b]
    \centering
    \includegraphics[width=\textwidth,trim={0cm 0 0 0},clip]{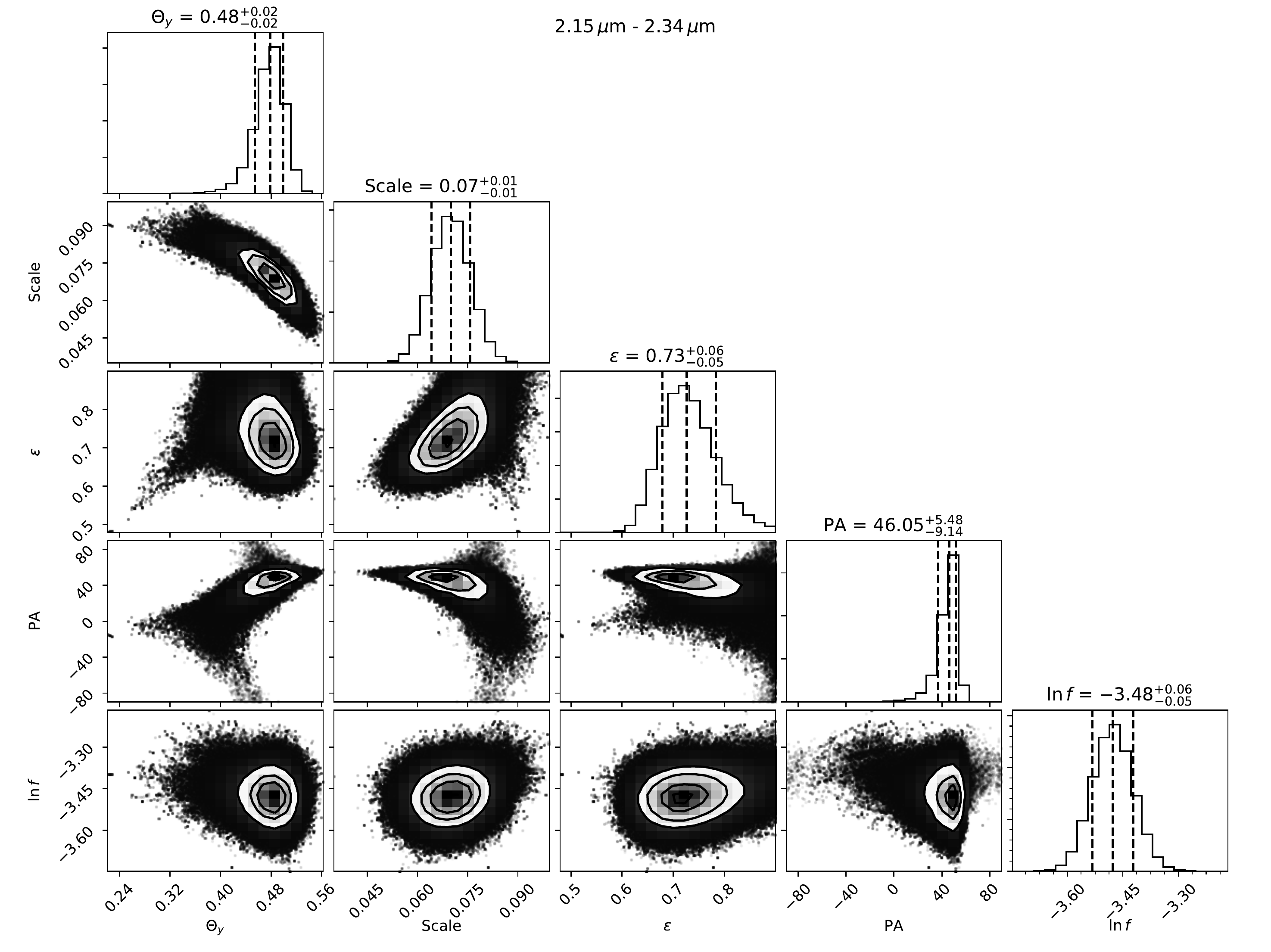}
    \caption{Corner plot \citep{foreman-mackey_corner.py:_2016} of the PDF of each parameter used in the elongated Gaussian fit in Section\,\ref{Section:angvis} for 2.15\,$\mu$m$-$2.34\,$\mu$m.}
    \label{fig:AppGaussAll}
\end{figure}

In Section\,\ref{Section:angvis} we fit the visibility squared with an elongated Gaussian model. Here we provide the corner plots of those fits in Figures\,\ref{fig:AppGauss206}$-$\ref{fig:AppGaussAll}.\\

{}

{}

{}

{}

{}

\clearpage

\section{Alternative Models}\label{sec:Appendix_models}

In Section\,\ref{sec:Grav_Mod}, we discussed the point source model and mentioned the use of more complex Gaussian models. Here we describe the Gaussian models. There are two Gaussian models we considered, a radially symmetric Gaussian model and an elongated Gaussian model. The radially symmetric Gaussian model has a central Gaussian with a fixed position and brightness, similar to the point source model; however, its FWHM was not fixed. Each non-fixed point has four free parameters: FWHM, $\Delta$RA position, $\Delta$Dec position, and relative log amplitude flux (defined as the log of a Gaussian component's amplitude relative to the fixed point). Unlike the point source model, the fixed amplitude was not set to 0.15 because it is the integrated flux of the Gaussian that would need to be set to this value. The integration adds computational expense. To minimise computational expense, we instead set the log amplitude of the fixed Gaussian to 10 and set the stipulation that the fixed point was the highest in amplitude. Each log amplitude was then set relative to the fixed point and the image was normalised to a summed flux of one in linear space. The bounds are given in Table\,\ref{tab:sym_gauss_bounds}. The radially symmetric Gaussian model has 4$\mathrm{N}_G$-3 free parameters, where $\mathrm{N}_G$ is the number of Gaussians and not including $s_f$, when calculating the BIC.

\begin{table}[h]
    \centering
    \begin{tabular}{l|cc}
        \hline
         Parameter&Lower&Upper  \\ \hline\hline
         $\Delta$RA&-10\,mas&10\,mas\\
         $\Delta$Dec&-10\,mas&10\,mas\\
         $A_f$&0&10\\
         FWHM&$10^{-4}$\,mas&5\,mas
    \end{tabular}
    \caption{The bounds for each free parameter of the radially symmetric Gaussian model. $A_f$ is the relative amplitude flux.}
    \label{tab:sym_gauss_bounds}
\end{table}{}

The elongated Gaussian model was identical to the radially symmetric Gaussian model except that the FWHM parameter was replaced with the major axis FWHM and two extra free parameters were introduced which are the major-minor axis ratio and the PA of the major axis. The bounds are in Table\,\ref{tab:elong_gauss_bounds}. The major axis FWHM and the axis ratio were free parameters for the fixed Gaussian. The chosen setup gives the model 6$\mathrm{N}_G$-3 free parameters when calculating the BIC. Both Gaussian models sampled 2 to 16 Gaussians.

\begin{table}[hb]
    \centering
    \begin{tabular}{l|cc}
        \hline
         Parameter&Lower&Upper  \\ \hline\hline
         $\Delta$RA&-10\,mas&10\,mas\\
         $\Delta$Dec&-10\,mas&10\,mas\\
         $A_f$&0&10\\
         FWHM$_\mathrm{major}$&$10^{-4}$\,mas&5\,mas\\
         $\epsilon$&0&1\\
         PA&$0^\circ$&$180^\circ$
    \end{tabular}
    \caption{The bounds for each free parameter of the elongated symmetric Gaussian model. $A_f$ is the relative amplitude flux, FWHM$_\mathrm{major}$ is the FWHM of the major axis, and $\epsilon$ is the axis ratio.}
    \label{tab:elong_gauss_bounds}
\end{table}{}

\end{document}